\documentclass[12pt]{article}
\usepackage[utf8]{inputenc}
\usepackage{textcomp} 
\usepackage{graphicx}
\usepackage{ulem}
\usepackage[dvipsnames]{xcolor}
\usepackage{amssymb}
\usepackage{amsmath}
\usepackage[a4paper, portrait, margin=1.5cm]{geometry}
\usepackage{caption}
\usepackage{wrapfig}
\usepackage{color}

\begin{document}


\title{Non-polaritonic effects in cavity-modified photochemistry}
\author{Philip A. Thomas$^1$\footnote{p.thomas2@exeter.ac.uk}, Wai Jue Tan$^1$, Vasyl G. Kravets$^2$,
\\
Alexander N. Grigorenko$^2$ and William L. Barnes$^1$\footnote{w.l.barnes@exeter.ac.uk} \\ \\
\small{$^1$Department of Physics and Astronomy, University of Exeter,} \\ \small{Exeter, EX4 4QL, United Kingdom} \\
\small{$^2$School of Physics and Astronomy, University of Manchester,} \\ \small{Manchester, M13 9PL, United Kingdom}}
\date{}

\maketitle

\begin{abstract}
Strong coupling of molecules to vacuum fields has been widely reported to lead to modified chemical properties such as reaction rates.
However, some recent attempts to reproduce infrared strong coupling results have not been successful, suggesting that factors other than strong coupling may sometimes be involved.
Here we re-examine the first of these vacuum-modified chemistry experiments in which changes to a molecular photoisomerisation process in the UV-vis spectral range were attributed to strong coupling of the molecules to visible light.
We observed significant variations in photoisomerisation rates consistent with the original work; however, we found no evidence that these changes need to be attributed to strong coupling.
Instead, we suggest that the photoisomerisation rates involved are most strongly influenced by the absorption of ultraviolet radiation in the cavity.
Our results indicate that care must be taken to rule out non-polaritonic effects before invoking strong coupling to explain any changes of chemical properties arising in cavity-based experiments.
\end{abstract}

\newpage
\noindent
Strong light-matter coupling occurs when the rate of photon exchange between an ensemble of molecular resonators and a confined electromagnetic mode exceeds losses to the environment~\cite{garcia2021manipulating, lidzey1998strong}.
This leads to the formation of two polariton modes, separated by an energy $\hbar\Omega$, that inherit properties from both the light and matter states~\cite{torma2014strong, ebbesen2016hybrid}.
Strong coupling has been investigated in fields such as lasing~\cite{ramezani2017plasmon}, Bose-Einstein condensation~\cite{hakala2018bose} and singular optics~\cite{thomas2022all}.
Perhaps the most exciting area of study is polaritonic (or vacuum-modified) chemistry~\cite{herrera2016cavity, feist2017polaritonic, herrera2020molecular, garcia2021manipulating, nagarajan2021chemistry, yuen2022polariton}: if, as hoped, strong coupling changes the energy levels of molecules and their coherent interaction, can it modify chemical processes?
It has been reported that strong coupling of electronic transitions in molecules to visible light can suppress photochemical processes~\cite{hutchison2012modifying, munkhbat2018suppression, peters2019effect, galego2016suppressing, mony2021photoisomerization}.
More attention has been given to the potential of strong coupling of vibrational modes in molecules to infrared radiation, known as vibrational strong coupling (VSC)~\cite{shalabney2015coherent, long2015coherent, george2015liquid}. VSC has been reported to modify chemical processes including reaction rates~\cite{thomas2016ground, lather2019cavity}, enzyme activity~\cite{vergauwe2019modification, lather2021improving} and reaction pathways~\cite{thomas2019tilting, sau2020modifying}.
While the idea of modifying chemical processes using light is not a new one~\cite{letokhov1973use, karlov1974laser, bard1979photoelectrochemistry, zhang2013plasmonic}, polaritonic chemistry is especially exciting because the light involved is not introduced by for example a laser beam, rather it is strong coupling to vacuum electromagnetic fields that is thought to be sufficient to modify chemical processes, even in the absence of incident light~\cite{hutchison2012modifying}.

The claims of polaritonic chemistry are compelling, but controversy remains~\cite{simpkins2021mode, wang2021roadmap}; a full understanding has yet to be developed.
Two recent attempts to reproduce VSC experiments were unsuccessful~\cite{imperatore2021reproducibility, wiesehan2021negligible} and some theoretical works suggest that VSC may not produce measurable changes to the chemical processes they considered~\cite{vurgaftman2020negligible, du2021nonequilibrium}.
Many VSC experiments have been performed on molecules in solution in a flow cell that doubles as an optical microcavity~\cite{nagarajan2021chemistry}; and it has been suggested that this experimental design could give rise to false positives, for example due to the diffusion of reactants or products, temperature shifts or slight changes in cavity thickness~\cite{simpkins2021mode}.

A major challenge in polaritonic chemistry is the design of suitable control experiments.
Strong coupling requires molecules to be placed in a structure (such as a planar microcavity) that possesses an electromagnetic mode tuned to match the energy of the molecular resonance of interest.
It has also been reported that VSC vacuum-modified chemistry requires the molecular resonance to coincide with the cavity's normal-incidence resonance (i.e. zero-detuning at normal incidence), even if the experiment is probed with light at an oblique angle~\cite{thomas2016ground, thomas2019tilting, garcia2021manipulating}. This is a challenging idea~\cite{li2022molecular}, that is gaining some theoretical support~\cite{vurgaftman2022comparative}, and is one that we explore for strong coupling to visible light later in this work.
Determining the effect of strong coupling on chemical processes requires one to compare this `on-resonance' structure with one where no strong coupling is present: either an `off-resonance' cavity~\cite{vergauwe2019modification, hirai2020modulation} (where the optical resonance does not match the molecular resonance of interest) or a `cavity-free'~\cite{peters2019effect, lather2021improving} structure (where any confined electromagnetic modes should be of negligible strength).
Some studies use a combination of `off-resonance' and `cavity-free' structures~\cite{hutchison2012modifying, thomas2016ground, thomas2019tilting, sau2020modifying}.
Both control approaches have limitations.
An appropriate `off-resonance' cavity control requires the optical resonance to be fully detuned from the molecular resonance and must not inadvertently have a higher- or lower-order cavity mode tuned to match some other molecular mode which could influence chemical processes.
Meanwhile, all so-called `cavity-free' structures (such as a planar microcavity that lacks a top mirror~\cite{hutchison2012modifying}) will still support some manner of confined electromagnetic mode~\cite{canales2020abundance}, many of which can lead to strong coupling~\cite{georgiou2020strong, thomas2021cavity}.
Furthermore, changing the structure will modify the intensity of light incident on the molecules, making direct comparison between the results from an `on-resonance' cavity and a `cavity-free' structure difficult, especially when studying photochemical processes~\cite{hutchison2012modifying, peters2019effect, mony2021photoisomerization}.
To be persuasive, polaritonic chemistry experiments might best compare the results from a series of structures ranging from ones where strong coupling is present to ones where no strong coupling is present.
This could be achieved by systematically increasing the thickness of a cavity (ideally covering a large, continuous range of thicknesses where strong coupling to different cavity modes is observed) or by studying cavities with different levels of field confinement (for example, by making cavities with mirrors of different thicknesses).
If strong coupling does influence the chemical process, one could then look for a correlation, for example by plotting reaction rate in a cavity against the cavity mode detuning. 
Very few experiments in the literature fully meet these criteria~\cite{ahn2022modification}.

A decade ago Hutchison \textit{et al.}~\cite{hutchison2012modifying}, in a pioneering experiment, studied the photoisomerisation of spyropyran (SPI) to merocyanin (MC), a reaction promoted by ultraviolet (UV) light. In particular, they examined the photoisomerisation under conditions for strong coupling to visible light.
This study has been credited with initiating the field of polaritonic chemistry~\cite{herrera2020molecular, garcia2021manipulating}.
Hutchison \textit{et al.} measured the rate of photoisomerisation in two planar microcavities and two `cavity-free' structures (i.e. in the absence of a top mirror), each of different thickness, concluding that strong coupling of MC to a visible-light cavity mode suppressed the rate of photoisomerisation.
In the work reported here we measured the photoisomersation rate of SPI in a range of planar structures for a large range of cavity thicknesses.
Our experimental results are broadly consistent with those reported by Hutchison \textit{et al.}:
we found that photoisomerisation rates can vary dramatically as cavity thickness and incident angle are changed.
However, our detailed analysis leads us to suggest that strong coupling of visible light to MC has no discernible effect on the measured photoisomerisation rate.
Instead, we suggest that changes in the photoisomerisation rate are better accounted for by the effectiveness of the cavity at absorbing the UV light necessary for photoisomerisation.
Our results lead us to suggest caution in interpreting the results of polaritonic chemistry experiments, in particular highlighting the need for vigilance in ruling out non-polaritonic effects.

\section*{Results}
\subsection*{Experiment}

Spiropyran (SPI) is a photochromic molecule that, when exposed to UV radiation (energy 3.5 eV), undergoes photoisomerisation to merocyanin (MC)~\cite{berkovic2000spiropyrans}.
Exposure to visible light can convert MC back into SPI.
The chemical structures of SPI and MC are given in Fig. \ref{fig:design}a and their extinction spectra are plotted in Fig. \ref{fig:design}b (optical constants in Supplementary Section S1.1): SPI is transparent in the visible whilst MC has a strong absorption peak at 2.2 eV.
Strong coupling to MC's molecular resonance at 2.2 eV is possible and the photosensitive nature of SPI/MC makes it an attractive molecule for use in strong coupling experiments~\cite{schwartz2011reversible, thomas2020new, thomas2021cavity, thomas2022all}.
The strength of light-matter coupling is directly proportional to the square root of the number density of coupled molecules~\cite{torma2014strong}.
UV irradiation of a SPI-filled cavity gradually increases the number of MC molecules in the cavity, allowing one to observe the transition from the weak to strong coupling regime.

We characterised the SPI-MC transition using spectroscopic ellipsometry (schematic in Fig. \ref{fig:design}c).
Ellipsometry measures the complex reflection ratio $\rho$ in terms of the parameters $\Psi$ and $\Delta$~\cite{tompkins2005handbook}:
\begin{align}
    \rho = \frac{r_p}{r_s} = \tan(\Psi)e^{i\Delta}.
\end{align}
$r_p$ and $r_s$ are the Fresnel reflection coefficients for p- and s-polarised light, respectively.
$\tan(\Psi)$ is the amplitude of $\rho$ and gives the ratio of the moduli of $r_p$ and $r_s$.
P-polarised modes appear as local minima in $\Psi$ ($|r_p|<|r_s|$ when $\Psi < 45^{\circ}$) and s-polarised modes appear as local maxima in $\Psi$ ($|r_s|<|r_p|$ when $\Psi > 45^{\circ}$).
$\Delta$ is the difference in the phase shifts undergone by p- and s-polarised light upon reflection.
Jumps in $\Delta$ correspond to photonic modes; points where $r_{p,s}=0$ are associated with phase singularities~\cite{thomas2022all}.
The light source of our ellipsometer, a Xe arc lamp (further details, including emission spectrum, in Supplementary Section S1.4), emits a small quantity of UV radiation, allowing us to measure the transition from the weak to strong coupling regime by undertaking a series of repeated measurements.
An example of this is shown in Figs. \ref{fig:design}d ($\Psi$) and \ref{fig:design}e ($\Delta$).
Ellipsometry is routinely used to determine the optical constants and thicknesses of thin films in multilayer structures.
A model is used to calculate $\Psi$ and $\Delta$ spectra and the best match to experimental $\Psi$ and $\Delta$ spectra is achieved by adjusting the optical constants and thicknesses of each film.
Therefore, one series of measurements for a cavity allows us to determine the thickness of the SPI/MC film and to track the increase in MC concentration as a function of time (and hence photoisomerisation rate - see Methods), thereby allowing us to monitor the conversion rate.

\begin{figure}[tb]
\includegraphics[scale=1]{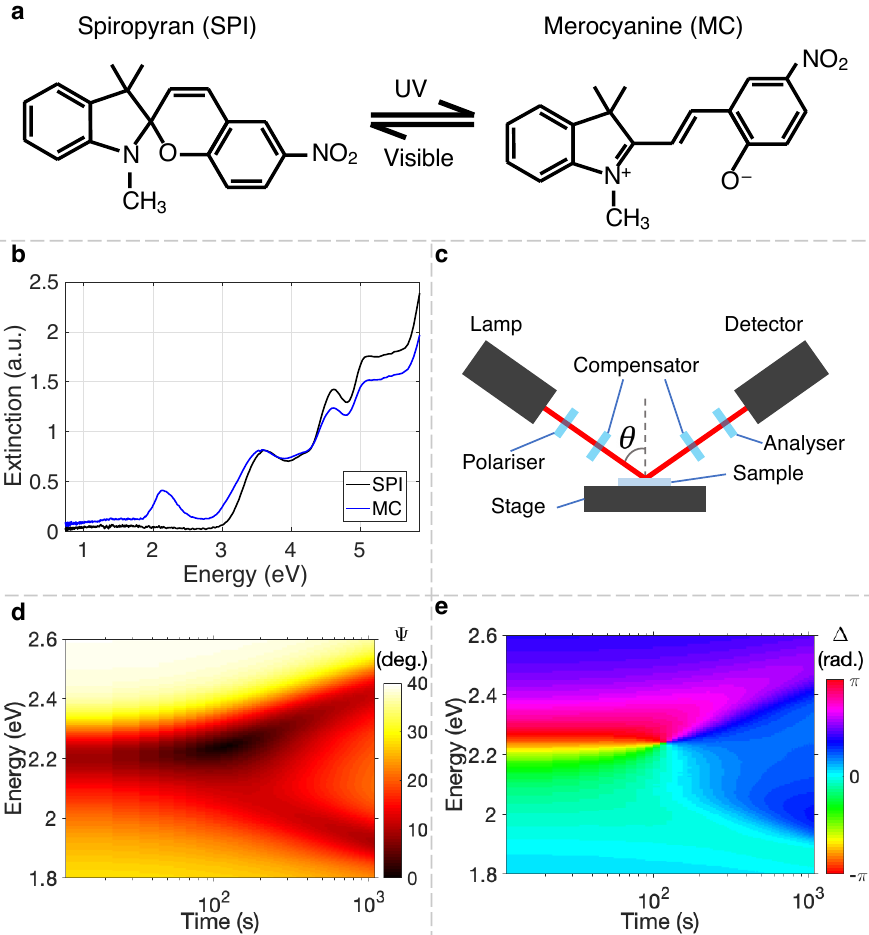}
\centering
\caption{\textbf{Strong coupling in photochromic organic films.}
(a) Chemical structures of spiropyran (SPI) and merocyanin (MC).
(b) Extinction spectra of SPI (black line) and MC (blue line) films normalised to extinction through a fused silica substrate.
(c) Ellipsometer schematic.
(d,e) Spectra of the ellipsometric parameters (d) $\Psi$ and (e) $\Delta$ of an Al (10 nm)/SPI (131 nm)/Al (10 nm) structure as a function of UV irradiation time acquired at an incident angle of $\theta=65^{\circ}$.
UV exposure increases the number of MC molecules in the SPI/MC film; after $\sim 100$ s the MC density is high enough to allow the cavity mode present in the structure to undergo strong coupling to the 2.2 eV resonance in MC, resulting in the formation of two polariton modes.
Panel (e) shows the associated creation of a phase singularity similar to what is reported in ref. \cite{thomas2022all}.}
\label{fig:design}
\end{figure}

We studied SPI-MC photoisomerisation in three types of structure (see Methods for fabrication details): planar microcavity (SPI/MC film between two thin metal films), mirror (SPI/MC film on a thin metal film), and bare substrate (SPI/MC film on a bare substrate).
In each case the range of SPI/MC film thickness was at least 100-400 nm.
This allowed us to measure the SPI-MC photoisomerisation rate as successive confined electromagnetic field modes were gradually tuned to match the MC molecular resonance at 2.2 eV.
We primarily used Al thin films since Al cavities can support high-quality cavity modes in the visible and UV.
We repeated our experiments with Ag and Cr films and reached the same conclusions; see Supplementary Sections S4 and S5.4.

We first studied photoisomerisation in the incident angle range $45^{\circ} \leq \theta \leq 75^{\circ}$ (the range of our ellipsometer).
We then studied Al cavities at normal incidence to determine whether reports of angle independence in VSC experiments (that reaction rate modification occurs most strongly in experiments designed to give zero detuning at normal incidence) might also apply for strong coupling of MC with visible light.

\subsection*{Probing photoisomerisation at oblique incident angles}

\begin{figure}[tbp]
\includegraphics[scale=1]{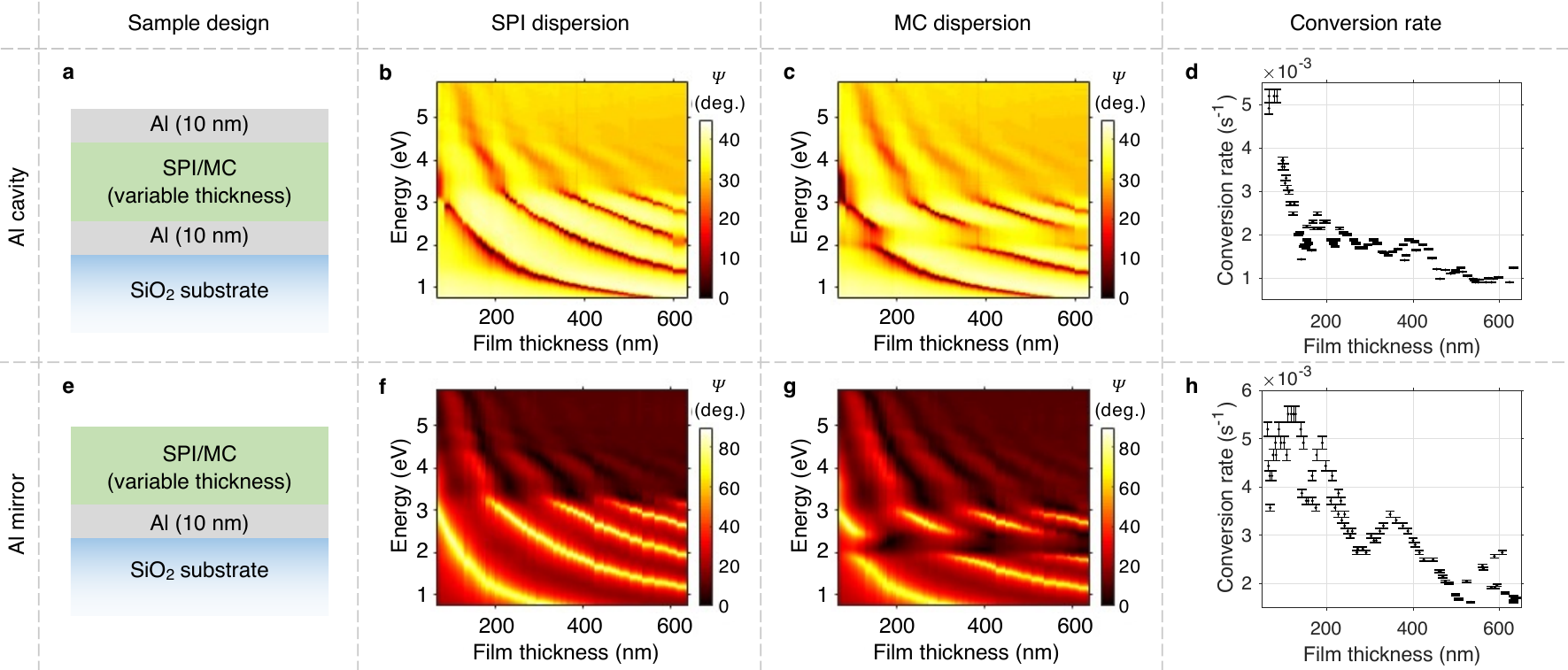}
\centering
\caption{\textbf{Photochemistry in Al structures.} (a,e) Sample designs, (b,d) SPI and (c,g) MC dispersion plots constructed from $\Psi$ and (d,h) photoisomerisation rates for (a-d) Al cavity and (e-h) Al mirror structures.
All data acquired at $\theta=65^{\circ}$.}
\label{fig:al-big}
\end{figure}

The results from our Al structures probed at $\theta=65^{\circ}$ are presented in Fig. \ref{fig:al-big}.
Sample designs are shown in Figs. \ref{fig:al-big}a,e.
In Figs. \ref{fig:al-big}b,c,f,g we show dispersion plots constructed from $\Psi$ spectra measured for structures with a range of SPI/MC film thicknesses.
SPI dispersion plots (panels b,f) were constructed from $\Psi$ spectra acquired after 11 seconds of UV exposure, when the cavities contained a negligible quantity of MC.
In the Al cavity (panel b) we observe a series of sharp cavity modes in the visible region of the spectrum, where SPI is transparent.
SPI becomes strongly absorbing above 3 eV (see Fig. \ref{fig:design}b), resulting in a deterioration of the cavity mode quality.
The dispersion plot for the Al mirror/SPI structure is shown in panel f.
This structure supports a series of leaky modes, alternating between transverse electric (maxima in $\Psi$) and transverse magnetic (minima in $\Psi$) modes.
These modes are of lower-quality than the cavity modes observed in the Al cavity structure in panel b; they are also sharpest in the visible and less well-defined in the UV.
The MC dispersion plots (panels c,g) were constructed from $\Psi$ spectra acquired after 1,100 seconds of UV exposure.
The MC molecular resonance at 2.2 eV strongly couples to the optical modes in the Al cavity (panel c) and the Al mirror (panel g) structures, giving clear anticrossing signatures.
The mirror structure is commonly used as a `cavity-free' control, but it is clear from the data that it too supports strong coupling.

In addition to the initial SPI and final MC spectra presented in Fig. \ref{fig:al-big}b,c,f,g, we acquired additional ellipsometry spectra for these samples every 11 s in the intervening time (see e.g. Figs. \ref{fig:design}d,e).
To investigate the change in MC concentration during UV exposure we first determined the optical constants of the SPI/MC film by fitting to ellipsometry spectra at each time; see Methods and Supplementary Section 2.1 for further details.
We then tracked the growth in MC concentration by plotting $\alpha(E_{\text{MC}})$, the absorption coefficient of the SPI/MC film at $E_{\text{MC}} = 2.2$ eV (which is directly proportional to MC concentration, see Supplementary Figure S5) as a function of time (examples in Supplementary Section S2.2).
A simple exponential function cannot be fit to these data.
To quantify the rate of conversion in a structure we introduce the conversion constant $\kappa$, the reciprocal of the time it takes for $\alpha(E_{\text{MC}})$ to reach 3.5 µm$^{-1}$ in that structure.
This target value was chosen because it is high enough to place the system well within the strong coupling regime (see Supplementary Section S2.1).
When plotting $\kappa$, error bars are calculated from the time interval between successive measurements (11 s).
In Figs. \ref{fig:al-big}d,h we plot the conversion constants for each structure as the SPI film thickness is varied.
Both plots show the same basic trend: $\kappa$ is largest in thinner films ($\sim5\times 10^{-3}$ s$^{-1}$ for thicknesses below 100 nm), levelling off at around $1\times 10^{-3}$ s$^{-1}$ for the thickest films.
SPI is lossy in the UV, so it is harder for UV to penetrate through thicker films.
Therefore, the optical constants of thicker films will change more slowly than they will for thinner films, resulting in lower conversion rates.
Each plot also shows smaller, approximately sinusoidal, deviations from this trend that become weaker for thicker films.
These trends are broadly consistent with Hutchison \textit{et al.}'s~\cite{hutchison2012modifying} results: an on-resonance cavity (thickness 155 nm in our case) has a lower reaction rate than both a mirror sample with the same film thickness (even when accounting for differences in UV intensity impinging on the sample) and a fully detuned (thinner) off-resonance cavity (see also Supplementary Section S2.2).
However, in contrast to Hutchison \textit{et al.}~\cite{hutchison2012modifying}, we found the trends in $\kappa$ were independent of the choice of target absorption coefficient, even if that target value was insufficient for strong coupling (see Supplementary Section S2.3).

We repeated the experiment for Al cavities probed at $\theta=45^{\circ}$, $55^{\circ}$ and $75^{\circ}$.
The plots of $\kappa$ against SPI/MC film thickness are given in Supplementary Section S3.
While a similar $\kappa$ trend occurs for all $\theta$, the trend is dispersive, i.e. the data show that the trend varies with $\theta$.
We also repeated this experiment at $\theta=65^{\circ}$ for a series of different structures: Ag cavity, Ag mirror, Cr mirror and SPI/MC films on bare substrates of Si and amorphous glass (results in Supplementary Section S4).
Data from all structures showed the same fundamental trend in photoisomerisation ($\kappa$ decreases with increasing SPI/MC film thickness).
The metal-based structures all show the same additional undulation as observed in Figs. \ref{fig:al-big}d,h; these undulations are not clear in the metal-free SPI/MC on Si and amorphous glass substrate structures, which we note provide far weaker field enhancements in the SPI/MC film than the metal-based structures.

\begin{figure}[!t]
\includegraphics[scale=1.0]{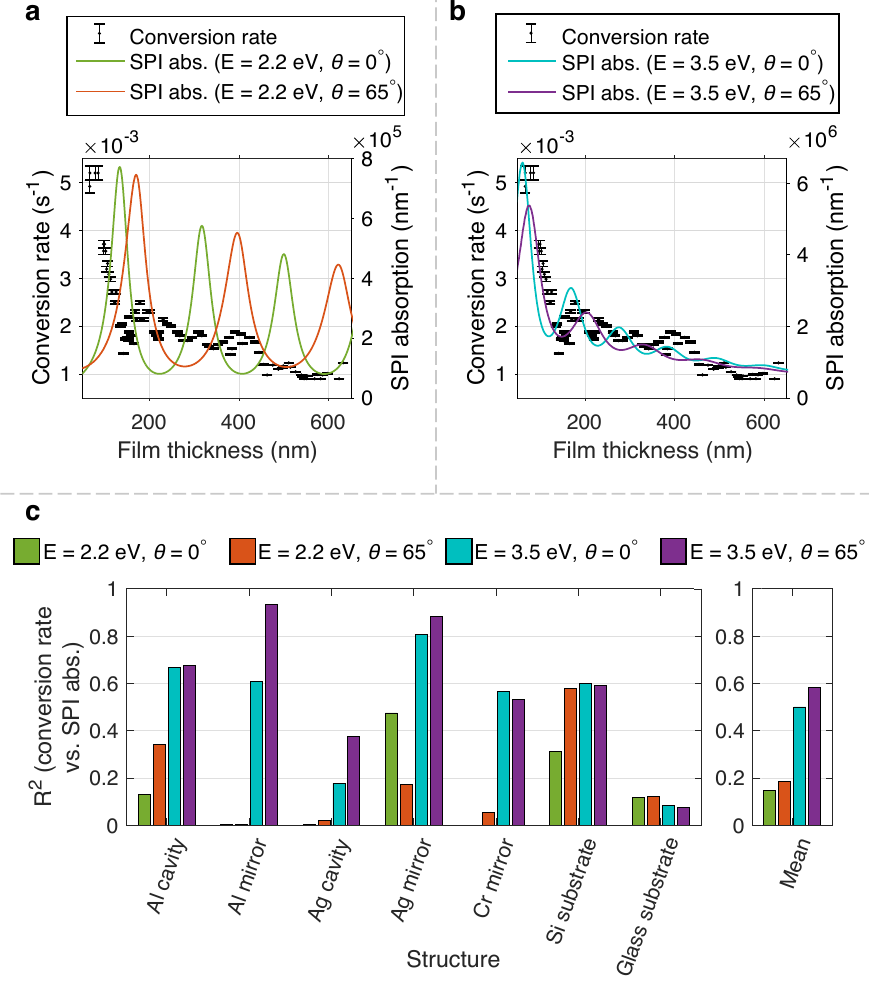}
\centering
\caption{\textbf{Accounting for photochemical reaction rates.} (a-b) Comparing experimentally determined SPI conversion rates in Al cavities (plotted in Fig. \ref{fig:al-big}d) with calculated SPI absorption normalised to SPI film thickness.
(a) The SPI absorption profiles calculated for $E=2.2$ eV incident at $\theta=0^{\circ}$ (light green line) and $\theta=65^{\circ}$ (dark green line) show no meaningful correlation with experimental data.
(b) The SPI absorption profiles calculated for $E=3.5$ eV incident at $\theta=0^{\circ}$ (light blue line) and $\theta=65^{\circ}$ (dark blue line) have much stronger correlations with experimental data, the best match occurring for $\theta=65^{\circ}$.
(c) Bar chart showing least-squared $R^2$ correlation coefficient between conversion rate profile (as measured at $\theta=65^{\circ}$) and calculated SPI absorption profiles for a range of different structures.
The mean values of $R^2$ for all structures are also plotted.
}
\label{fig:abs}
\end{figure}

\subsection*{Determining the cause of observed photoisomerisation trends}

The data in Figs. \ref{fig:al-big}d,h show that $\kappa$ varies with SPI/MC film thickness, but what causes these trends?
Hutchison \textit{et al.}~\cite{hutchison2012modifying} suggested that strong coupling of optical cavity modes to the MC resonance at 2.2 eV can lead to a reduction of the rate $\kappa$.
If so, we might expect a correlation between the absorption of light of energy $E = 2.2$ eV in the SPI film and $\kappa$.
It has also been reported that VSC vacuum-modified chemistry requires the molecular resonance to coincide with the cavity's normal-incidence resonance, even if the experiment is probed with light at an oblique angle~\cite{thomas2016ground, thomas2019tilting, garcia2021manipulating, vurgaftman2022comparative}.
If this is also true when excitonic resonances (rather than vibrational resonances) of molecules are strongly coupled to visible (rather than infrared) light, then we might anticipate a correlation between $\kappa$ and the absorption of light of energy $E = 2.2$ eV in the SPI film at normal incidence.
It would also suggest that, as $\theta$ is varied, some features in the $\kappa$ profile should remain unchanged.
These expectations are not met: our results show the dispersive nature of $\kappa$ (Supplementary Section S3).
In Fig. \ref{fig:abs}a we re-plot the Al cavity $\kappa$ data in Fig. \ref{fig:al-big}d along with the calculated light absorbed in the SPI film (normalised to film thickness) for $E = 2.2$ eV light at incident angles of $\theta=0^{\circ}$ (green line) and $\theta=65^{\circ}$ (orange line).
(The amount of light absorbed in the SPI film is influenced by interference effects caused by the thickness of the SPI film and the impedance mismatch at each interface: it is not the same as the intensity of light impinging on the SPI film; see Supplementary Section S5.1 for further details.)
For our results to be consistent with Hutchison \textit{et al.}'s report that strong coupling suppresses photoisomerisation rates, we would expect the peaks in the SPI absorption profile at 2.2 eV to coincide with minima in the $\kappa$ profile.
However, from the data it is not clear that this is the case for either the $\theta=0^{\circ}$ or $\theta=65^{\circ}$ SPI absorption profile at 2.2 eV.

As an alternative, we consider the influence of the UV intensity inside the cavity on $\kappa$.
In Fig. \ref{fig:abs}b we plot the absorption profiles for $E = 3.5$ eV (the energy of SPI's lowest-energy UV absorption peak) at incident angles of $\theta=0^{\circ}$ (blue line) and $\theta=65^{\circ}$ (purple line).
These both give a much stronger match to $\kappa$ than 2.2 eV: the $\theta=65^{\circ}$ profile gives the best fit ($\theta=0^{\circ}$ gives an additional peak at 280 nm that does not match with any feature in the $\kappa$ data).
This suggests that the profile of $\kappa$ in Al cavities when illuminated at $\theta=65^{\circ}$ can be predicted by how effectively the cavity absorbs UV radiation.
In Supplementary Section S5.2 we show that the SPI absorption profile at 3.5 eV is a significantly better predictor of $\kappa$ than the SPI absorption profile at 4.6 eV (corresponding to another peak in SPI absorption, see Fig. \ref{fig:design}b).
In Supplementary Section S5.4 we repeated this analysis for all structures studied at $\theta=65^{\circ}$; the results are summarised in Fig. \ref{fig:abs}c, where we plot the least-squares $R^2$ correlation coefficient between $\kappa$ and the calculated absorption profiles (some limitations of these summary statistics are outlined in Supplementary Sections 5.3).
In every case, the best predictor of $\kappa$ is not the SPI absorption at 2.2 eV but the SPI absorption at 3.5 eV.
In no case is the $\theta=0^{\circ}$ profile a better match than the $\theta=65^{\circ}$ profile.

\begin{figure}[t]
\includegraphics[scale=1.0]{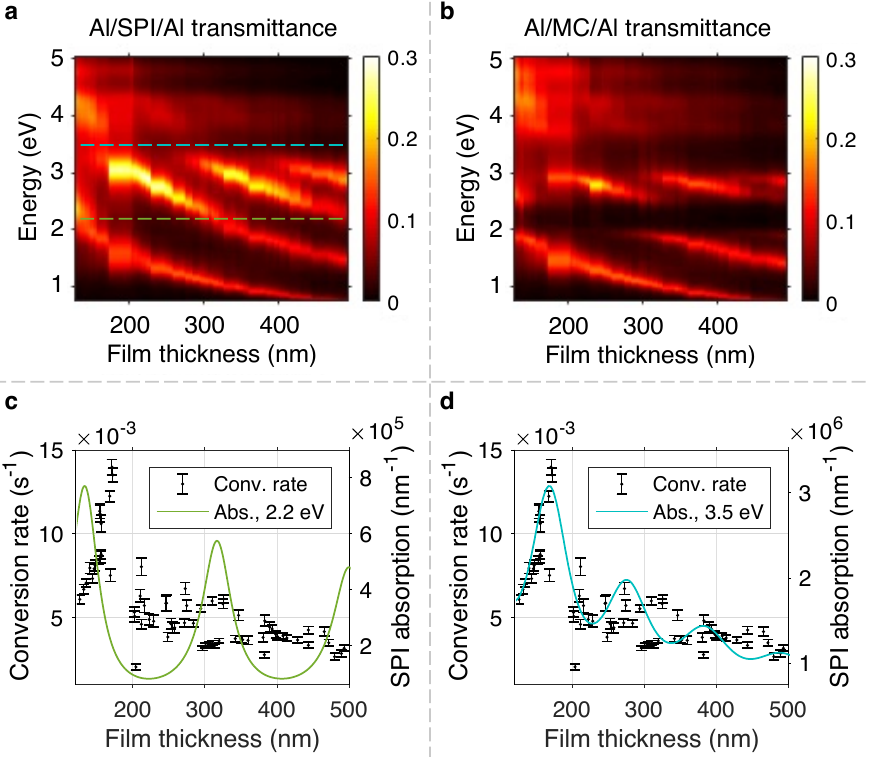}
\centering
\caption{\textbf{Photochemistry in Al cavities probed at normal incidence.} (a-b) Dispersion plots for (a) SPI and (b) MC constructed from transmission intensity spectra for a range of cavity thicknesses.
(c-d) Photoisomerisation rates for a range of Al cavity thicknesses (black data points), plotted with SPI absorption profiles for (c) $E=2.2$ eV, $\theta=0^{\circ}$ (green line) and (d) $E=3.50$ eV, $\theta=0^{\circ}$ (blue line).
The dashed green and blue lines in panel (a) show the energies at which the absorption profiles were calculated in panels (c) and (d) respectively.
}
\label{fig:norm}
\end{figure}

\subsection*{Cavity-modified photochemistry at normal incidence}

Several VSC experiments have found that vacuum-modified chemistry effects are strongest at zero detuning, i.e. when the molecular resonance energy matches that of the cavity mode at normal incidence~\cite{thomas2016ground, thomas2019tilting, garcia2021manipulating}. This observation is now gaining theoretical support~\cite{li2022molecular, vurgaftman2022comparative}.
To the best of our knowledge no comparable effect has been reported (either experimentally or theoretically) for strong coupling to visible light. To determine if a similar effect occurs with visible light, we studied photoisomerisation of Al microcavities at normal incidence; our results are presented in Fig. \ref{fig:norm}.
The SPI dispersion plot (Fig. \ref{fig:norm}a) reveals the same set of cavity modes (albeit slightly shifted in energy) as seen in Fig. \ref{fig:al-big}b.
Unlike the situation at $\theta=65^{\circ}$, at $\theta=0^{\circ}$ there is clear anticrossing taking place at $E=3.5$ eV, suggesting that the electric field confinement of Al cavity modes at normal incidence is high enough to allow them to strongly couple to the 3.5 eV SPI absorption peak.
The MC dispersion plot (Fig. \ref{fig:norm}b) shows, as expected, clear anticrossing at $E=2.2$ eV with Rabi splitting $\hbar\Omega \sim 700$ meV, consistent with previously reported results~\cite{schwartz2011reversible, hutchison2012modifying}.
Hutchison \textit{et al.} compared the reaction rate under coupling to the first-order cavity mode with the reaction rate in a thinner, off-resonance cavity; however, calculations in Supplementary Section S6 indicate that some splitting will be observed in all cavities likely to be made using spin-coating, leaving a question mark over such an off-resonance control.
In Fig. \ref{fig:norm}c, we plot profiles of the measured rate $\kappa$ and the calculated SPI absorption at 2.2 eV.
For our results to be consistent with Hutchison \textit{et al.}'s findings, then - as noted above - we would expect to see minima in $\kappa$ matching with peaks in the SPI absorption at 2.2 eV.
While we do see an increase in $\kappa$ between the first- and second-order cavity mode coupling conditions, this increase is a poor match with the decrease in SPI absorption; the trend is less decisive at greater thicknesses.
Instead, we find (Fig. \ref{fig:norm}d) that the UV absorption of SPI at 3.5 eV is a better predictor of $\kappa$.
This suggests, once again, that strong coupling to MC's 2.2 eV resonance has little if any measurable effect on $\kappa$.

These results do not challenge the observation of the normal-incidence effect in VSC experiments, but they do appear to rule out any analogous effects in the cavity-modified SPI-MC photoisomerisation process.

\section*{Discussion}
Our experiments are not a like-for-like reproduction of Hutchison \textit{et al.}'s~\cite{hutchison2012modifying} pioneering work~\cite{herrera2020molecular, garcia2021manipulating}.
In particular, the UV illumination conditions in the two works are quite different.
Nevertheless, the experimental results from both studies are broadly in agreement.
However, the more extensive nature of our study leads us to offer an alternative explanation to that of strong coupling to MC's 2.2 eV resonance influencing SPI/MC photoisomerisation.
We suggest that photoisomerisation rates can be increased by enhancing the cavity's UV absorption.
We substantially varied our irradiation conditions by studying a range of structures including Al cavities, Ag cavities, and samples with no top mirrors.
By taking account of the way the UV light is absorbed in each structure we showed that all of our results can be explained by the same mechanism.
Since the rate of a photochemical reaction is the rate of photons absorbed by the reactant multiplied by the quantum yield of the reaction (i.e. the number of molecules that complete the reaction per absorbed photon), our results suggest that strong light-matter coupling with MC does not affect the quantum yield of the SPI-MC photoisomerisation process.
In Supplementary Section S7 we normalise our reaction rate data to SPI absorption at 3.5 eV (shown in Figs. \ref{fig:abs}-\ref{fig:norm}) and find no discernible trend.
We did not perform transient absorption experiments as done by Hutchison \textit{et al.}~\cite{hutchison2012modifying} since
their transient absorption measurements do not appear to contradict our own findings.
It has recently been argued that transient absorption experiments are particularly prone to misleading optical signatures, and may often have their origin in photoexcitation effects other than the generation of polariton states~\cite{renken2021untargeted}.
Our results do not immediately challenge other reports of strong coupling modifying photochemistry~\cite{munkhbat2018suppression, peters2019effect, mony2021photoisomerization} (especially those where light strongly couples to the reactant rather than the product).
However, these experiments may benefit from similar re-examinations to verify that changes in reaction rates are indeed due to strong coupling rather than the inadvertent enhanced absorption of photons critical for the photochemical process,
or some other effect. Our findings may provide an explanation for the recent work of Zeng \textit{et al.}~\cite{zeng2023control}, who also report that SPI/MC photoisomerisation is suppressed when the strong coupling condition is met, but find that fulgicide photoswitching is accelerated under strong coupling.
This could simply be explained by the two processes being promoted by absorption of different energy light, resulting in cavity-modified photochemistry being predicted by different absorption profiles.
In the SPI/MC case the off-resonance cavity more efficiently absorbs UV light than the on-resonance sample; in the fulgicide case the opposite might be true.

Our results have other important implications.
There is a growing appreciation that planar cavities filled with organic molecules are much more complicated than previously thought~\cite{renken2021untargeted, cheng2022molecular}, making systematic studies of strong coupling all the more essential.
In VSC, there are concerns around reproducibility~\cite{imperatore2021reproducibility, wiesehan2021negligible} and the limited agreement between experiment and theory~\cite{simpkins2021mode, wang2021roadmap}.
We have shown that what are sometimes called `cavity-free' controls can in fact support polaritonic states (Fig. \ref{fig:al-big}e-h, Supplementary Section S4) and that it can be very difficult to fully de-tune cavity and molecular modes (see Supplementary Sections S2.2 and S6, which show evidence of strong coupling to MC in highly detuned cavities).
Many reports about the effect of vibrational strong coupling on chemical processes have involved comparison of the results from on- and off-resonance samples alongside a cavity-free sample~\cite{thomas2016ground, thomas2019tilting, hirai2020modulation, sau2020modifying, kadyan2021boosting, joseph2021supramolecular, sandeep2022manipulating}.
These experiments are unlikely to provide a complete picture of the rich variety of optical and chemical processes that occur under the conditions for VSC.
It would be helpful if future studies explored a far greater range of cavity thicknesses, so as to help rule out non-polaritonic effects~\cite{mazin2022inverse}.
While our results concern cavity-modified photochemistry, we hope that our work provides a potential blueprint for future VSC studies that may help to resolve the current differences between experiment and theory.

In conclusion, we have re-examined the SPI-MC photoisomerisation process inside a microcavity, as originally investigated by Hutchison \textit{et al.}~\cite{hutchison2012modifying}.
We have shown that photoisomerisation rates can be modified by cavity-based changes in UV absorption in the SPI/MC film.
We found no evidence that strong coupling to MC's 2.2 eV molecular resonance needs to be invoked to explain the observed effect on photoisomerisation.
Our results suggest that care must be taken to account for the influence of non-polaritonic effects in polaritonic chemistry experiments.

\section*{Methods}
\subsection*{Sample fabrication}
SPI (1,3,3-trimethylindolino-6'-nitrobenzopyrylospiran) / PMMA poly(methyl-methacrylate) films were prepared by dissolving PMMA (molar weight 996 kDa) in toluene (1 wt$\%$ PMMA). 
SPI and PMMA powders were purchased from Sigma Aldrich.
SPI was then dissolved in the PMMA-toluene solution with a weight ratio of 3:2 SPI to PMMA.
The solution was then filtered using a syringe filter of pore size 0.2 µm.
SPI films were deposited by drop-casting a small amount of the solution $\sim 2$ µL slightly off-centre on the sample, and spun at a wide range of spin speeds, from 1500 rpm (revolutions per minute) to 9000 rpm, to ensure the greatest range of film thicknesses.
Ag, Al and Cr films were deposited by thermal evaporation.
Details of Al thin film evaporation are given in Supplementary Section S1.3.
Representative pictures of final samples are provided in Supplementary Section S1.2.
Unlike in our work, Hutchison \textit{et al.}\cite{hutchison2012modifying} added dielectric spacer layers between their SPI and metal films out of concern that the metal might interfere with the photoisomerisation process. In Supplementary Section S1.5 we repeated our experiments with spacer layers and found that including spacers did not alter our conclusions.

\subsection*{Determination of photoisomerisation rate using ellipsometry}
Spectroscopic ellipsometry at oblique angles was carried out using a J. A. Woollam Co. M-2000XI.
The measurement of a structure with a given SPI film thickness consisted of the collection of 100 ellipsometry spectra (one every 11 seconds).
Over the course of this measurement the SPI film was converted to MC by UV radiation from the ellipsometer's Xe arc lamp.
The thicknesses of the metal and SPI films were determined from the first measurement (assumed to be 100$\%$ SPI).
The MC molecular resonance at 2.2 eV was modelled using a Tauc-Lorentz oscillator and the optical constants for the SPI/MC film for each of the 100 measurements was determined by using the Tauc-Lorentz oscillator amplitude as the fit parameter.
The change in MC concentration was monitored by tracking how the SPI/MC film's absorption coefficient at 2.2 eV (which was linearly proportional to the MC concentration in the film) changed with time.
The plotted conversion rates, $\kappa$, were determined by measuring the time taken for the SPI/MC film to reach a target absorption coefficient at 2.2 eV.
For $\theta=65^{\circ}$, (Figs. \ref{fig:al-big}-\ref{fig:abs}) the target was 3.5 µm$^{-1}$; for $\theta=0^{\circ}$, the target was 5.8 µm$^{-1}$.
Further details can be found in Supplementary Section S2.
Transmission measurements were conducted using a modified J. A. Woollam Co. M-2000F ellipsometer.
The same procedure was used to determine $\kappa$, performing fits to transmission intensity spectra instead of the ellipsometric parameters $\Psi$ and $\Delta$.


\section*{Acknowledgements}
P.A.T. and W.L.B. acknowledge the support of the European Research Council through Project Photmat (ERC-2016-AdG-742222: www.photmat.eu).
W. J. T. acknowledges financial support from the Engineering and Physical Sciences Research Council (EPSRC) of the United Kingdom via the EPSRC Centre for Doctoral Training in Metamaterials (EP/L015331/1).
V.G.K. and A.N.G. acknowledge the support of Graphene Flagship programme, Core 3 (881603).
We thank Kishan S. Menghrajani for discussions in the preliminary stages of this work.
For the purpose of open access, the authors have applied a Creative Commons Attribution (CC BY) licence to any Author Accepted Manuscript version arising.

\section*{Author contributions}
W. L. B. and P. A. T. designed the experiment. W. J. T. fabricated samples and performed calculations.
P. A. T. and V. G. K. performed measurements.
V. G. K. and A. N. G. designed and built the modified ellipsometer setup used for normal incidence measurements in Fig. \ref{fig:norm}.
All authors contributed to discussions.
P. A. T. wrote the manuscript with input from all authors.

\section*{Competing interests}
The authors declare no competing interests.

\newpage
\begin{center}
    {\LARGE  Supplementary Information for:}
    \\ \vspace{0.5cm}
    {\LARGE Non-polaritonic effects in cavity-modified photochemistry}
    \\ \vspace{0.5cm}
    Philip A. Thomas$^1$, Wai Jue Tan$^1$, Vasyl G. Kravets$^2$,
\\
Alexander N. Grigorenko$^2$ and William L. Barnes$^1$ \\
\small{$^1$Department of Physics and Astronomy, University of Exeter,} \\ \small{Exeter, EX4 4QL, United Kingdom} \\
\small{$^2$School of Physics and Astronomy, University of Manchester,} \\ \small{Manchester, M13 9PL, United Kingdom}
\end{center}

\tableofcontents

\newpage
\addcontentsline{toc}{section}{S1. Further fabrication and sample details}
\section*{S1. Further fabrication and sample details}

\addcontentsline{toc}{subsection}{S1.1 SPI/MC optical constants}
\subsection*{S1.1 SPI/MC optical constants}
\renewcommand{\thefigure}{S1}
\begin{figure}[!h]
\includegraphics[width=17cm]{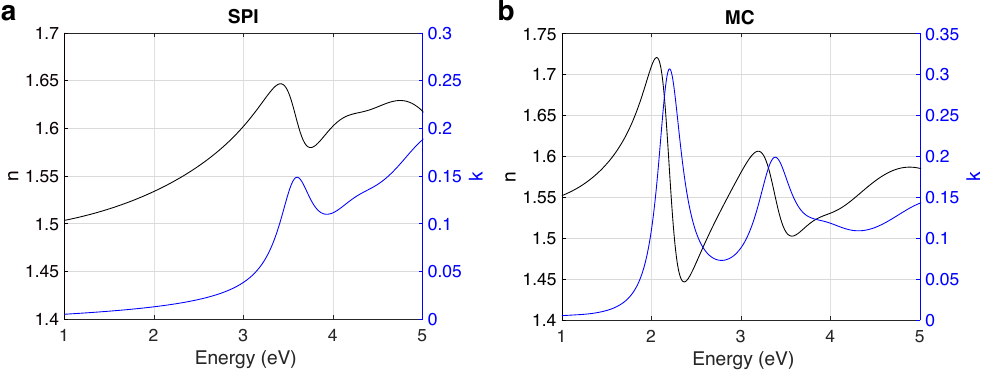}
\centering
\caption{Optical constants of (a) SPI and (b) MC films determined using spectroscopic ellipsometry.
}
\end{figure}

\addcontentsline{toc}{subsection}{S1.2 Sample images}
\subsection*{S1.2 Sample images}

\renewcommand{\thefigure}{S2}
\begin{figure}[!h]
\includegraphics[width=17cm]{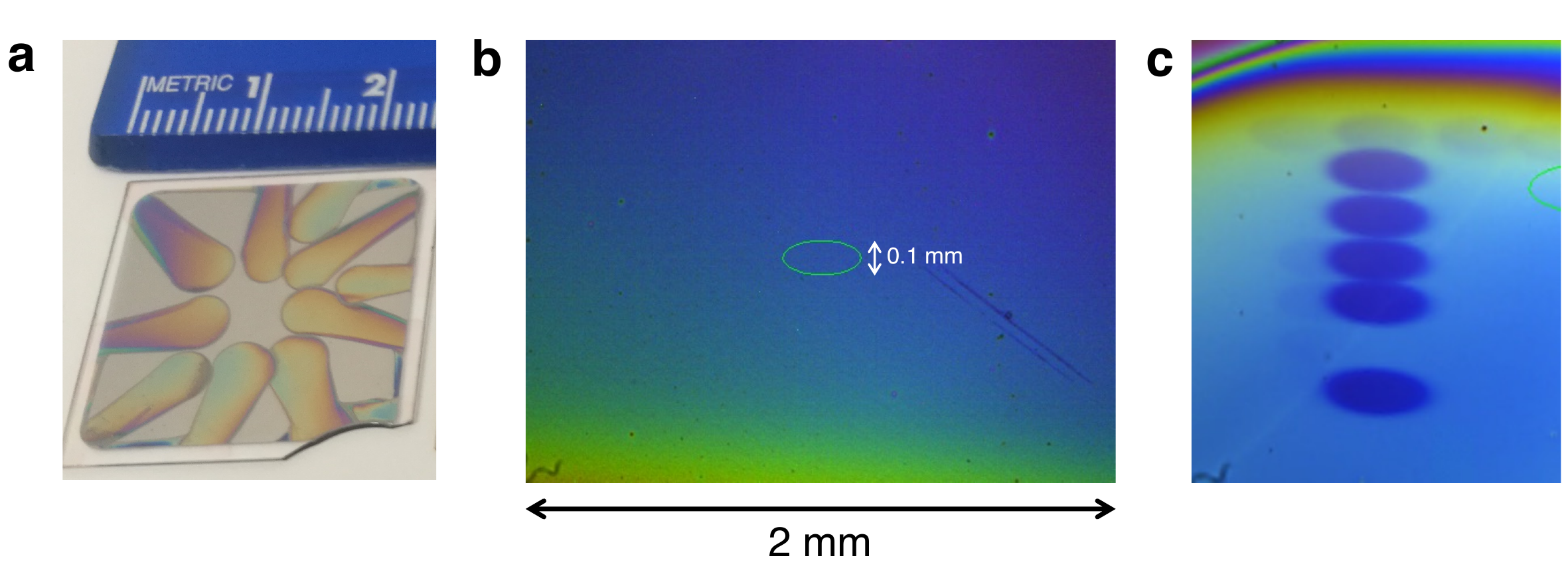}
\centering
\caption{(a) Representative picture of a sample used in our experiments. SPI was spin-coated off-centre to create a large range of SPI film thicknesses on a single substrate. (b) Representative close-up image of the surface of the sample. The green ellipse indicates the sample area interrogated in a single measurement at $\theta=65^{\circ}$. SPI films had effectively uniform thickness in a single measured region. The thickness of each measured spot was determined using ellipsometry (typical accuracy $\sim 1$ nm).
This allowed us to effectively create hundreds of cavities with a huge range of thicknesses (at least 100-400 nm) on a single substrate.
(c) Exposed regions were clearly distinguishable from unexposed regions, eliminating the risk of the same region being used for two different measurements.}
\label{fig:wafer}
\end{figure}

\newpage
\addcontentsline{toc}{subsection}{S1.3 Deposition of Al thin films using thermal evaporation}
\subsection*{S1.3 Deposition of Al thin films using thermal evaporation}
In thermal evaporation of thin metal films, metal pellets are placed in an evacuated chamber and heated with an electric current that is gradually increased until the metal begins to evaporate.
A thin film of metal is then gradually deposited on the target substrate also inside the vacuum chamber.
The target substrate is covered with a shield which can be removed once the metal deposition rate has stabilised. 
The deposition of high-quality thin films requires a slow and steady deposition rate.

The Al pellets typically used for thermal evaporation are covered in a thin oxide layer with a melting point far greater than that of pure Al.
Therefore, when heating the Al pellets for thermal evaporation, the Al inside the pellet will melt while the outer oxide layer remains intact.
Once the oxide layer is broken, all the vapour pressure built up inside the pellet is released, causing a sudden increase in chamber pressure.
This means that the Al deposition rate can rapidly increase, making it hard to control the thickness of the Al thin film.

To mitigate this effect we employed the following procedure:
\begin{enumerate}
    \item Maintain current at a lower level (before Al starts being deposited) to raise the Al vapour pressure inside pellets.
    \item If no deposition happens, gradually increase the current in small increments.
    \item When the outer oxide layer is broken, the pressure in the chamber suddenly increases.
    \item Lower the current to obtain a steady deposition rate of $0.1\sim 0.2$ nm s$^{-1}$.
    \item Remove the sample shield to deposit the metal film on the target substrate.
\end{enumerate}

\addcontentsline{toc}{subsection}{S1.4 Characterisation of UV light source}
\subsection*{S1.4 Characterisation of UV light source}

\renewcommand{\thefigure}{S3}
\begin{figure}[!h]
\includegraphics[scale=1.1]{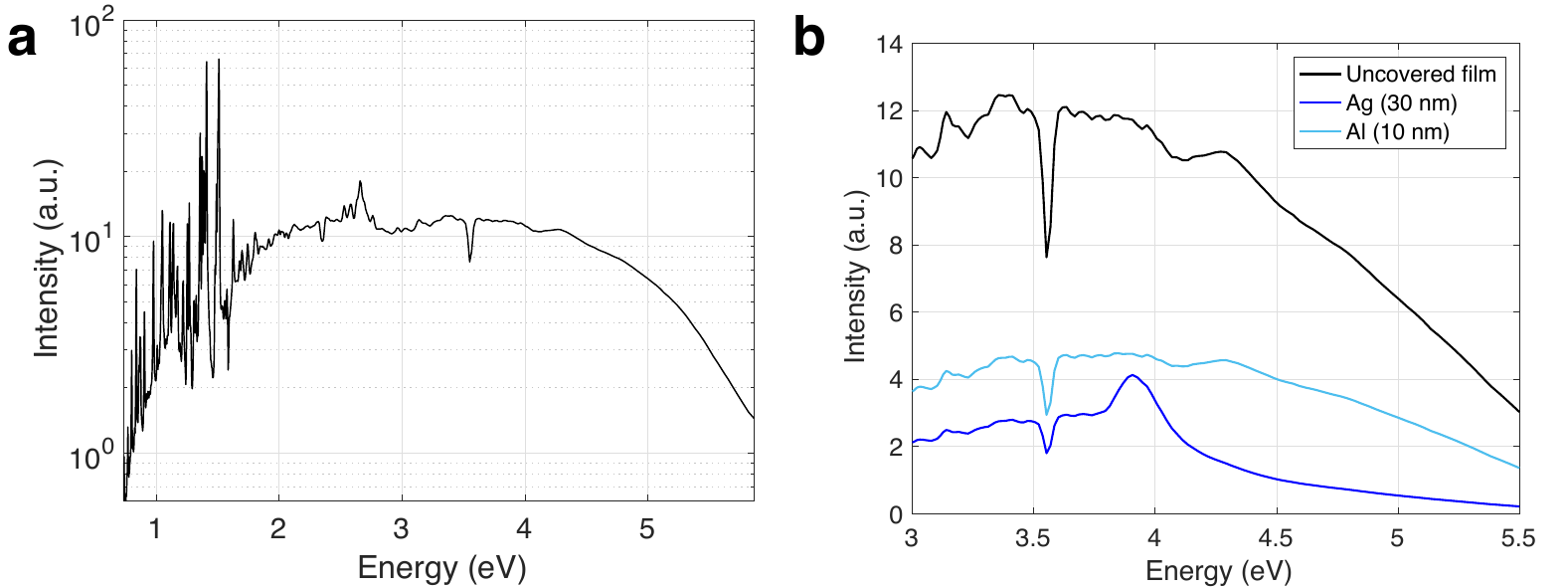}
\centering
\caption{(a) Intensity spectrum from ellipsometer's Xe arc lamp used to conduct experiments.
The lamp has a smooth, continuous emission spectrum in the ultraviolet.
(b) UV intensity spectrum of light from Xe arc lamp passing through air (black line), after transmission through a 30 nm Ag film (dark blue line) and after transmission through a 10 nm Al film (light blue line) at an incident angle of $\theta=65^{\circ}$.
The intensity profiles of UV radiation exposed to samples in various experiments (with Al top mirror, Ag top mirror, or no top mirror) are therefore very different, yet the results from all these experiments can be explained by the same mechanism. This suggests that the SPI$\rightarrow$MC photoisomerisation process dominates in all our experiments.
}
\end{figure}

\newpage
\addcontentsline{toc}{subsection}{S1.5 Photoisomerisation in cavities with PVA spacer layers}
\subsection*{S1.5 Photoisomerisation in cavities with PVA spacer layers}

\renewcommand{\thefigure}{S4}
\begin{figure}[!h]
\includegraphics[scale=1.2]{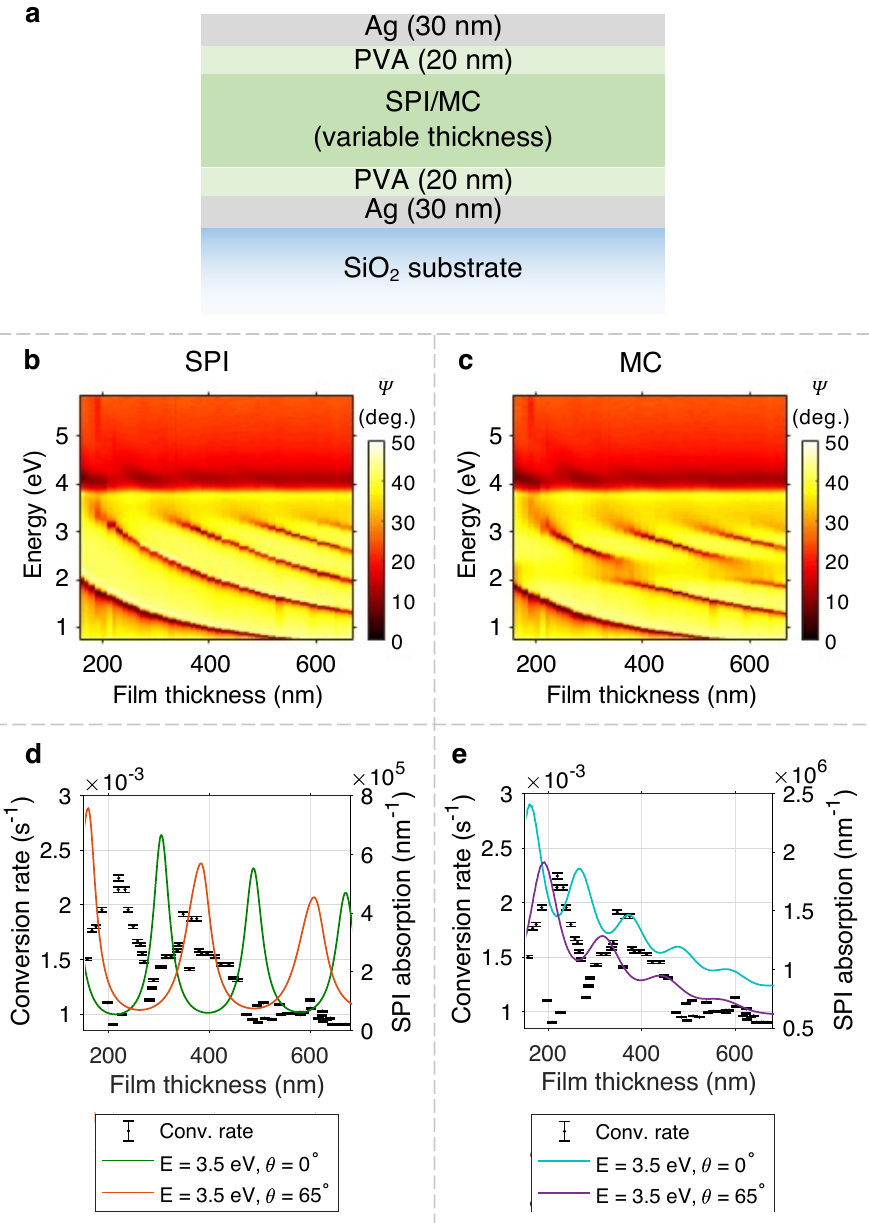}
\centering
\caption{Hutchison \textit{et al.}~\cite{hutchison2012modifyingSI} suggested that Ag evaporated directly on to the SPI film could influence the photochemical dynamics of SPI, leading the authors to produce cavities with PVA spacer layers separating the Ag films and SPI layer.
Here we repeat our experiment using this sample design, schematic in (a).
We observe the same (b) SPI and (c) MC dispersion plots as before (c.f. Supplementary Figure S(b,c)).
In (d) and (e) we observe the same trends in photoisomerisation rates as we did in previous experiments (c.f. Supplementary Figure S11(e,f)), and reached the same conclusion (that photoisomerisation rates are influenced by the cavity's UV absorption rather than by satisfying the MC strong coupling condition).
We conclude that the direct deposition of Ag films on the SPI film has a negligible influence on SPI/MC photochemistry, we conclude that the inclusion of PVA spacer layers is thus unnecessary in these experiments.}
\end{figure}

\newpage
\addcontentsline{toc}{section}{S2. Quantifying photoisomerisation rates in Al cavities}
\section*{S2. Quantifying photoisomerisation rates in Al cavities}

\addcontentsline{toc}{subsection}{S2.1 Modelling absorption in SPI/MC films}
\subsection*{S2.1 Modelling absorption in SPI/MC films}

\noindent SPI is transparent in the visible spectrum while MC has an absorption peak at $E_{\text{MC}}=2.2$ eV.
We tracked the rate of photoisomerisation by studying how the absorption coefficient at $E_{\text{MC}}$, $\alpha(E_{\text{MC}})$, changes under UV irradiation.
A higher value of $\alpha(E_{\text{MC}})$ corresponds to a greater MC concentration.
We determined the optical constants ($n,k$) of the SPI/MC film by fitting to data acquired using spectroscopic ellipsometry, an experimental technique expressly designed to accurately determine the optical constants of thin films.
The absorption coefficient $\alpha= 4\pi k/\lambda$ is then calculated from $k$, the imaginary component of the optical constants ($\alpha(E_{\text{MC}}) = 4\pi k(E_{\text{MC}})/\lambda_{\text{MC}}$).
We performed fits in the visible range (480 nm $<\lambda<$ 680 nm), where SPI is essentially transparent but MC has a molecular resonance at 2.2 eV.
Such absorption peaks are frequently modelled with simple Lorentz oscillators; however, we found that the MC molecular resonance was better modelled with a Tauc-Lorentz oscillator\cite{jelly}, a modified Lorentz oscillator that is used to model absorption in amorphous materials with zero absorption below a defined bandgap energy.
This was achieved by adding the following term to the dielectric function for MC:
\begin{align*}
    \epsilon_{TL} = \epsilon_{1} + i\epsilon_{2},
\end{align*}
where
\begin{align*}
    \epsilon_{1} &= \frac{2}{\pi}P\int_{Eg}^\infty \frac{\xi \epsilon_{2}(\xi)}{\xi^2 - E^2} \text{d}\xi
 \end{align*}
and
\begin{align*}   
    \epsilon_{2} &= \left[ \frac{AE_o\gamma(E-E_g)^2}{\left( E^2 - E_o^2\right)^2 + \gamma^2 E^2} \cdot \frac{1}{E} \right]
\end{align*}
for $E>E_g$, else $\epsilon_{2} = 0$.
$A$ (units eV) is the amplitude, $\gamma=0.315$ eV is the full width at half maximum, $E_o=2.175$ eV is the energy of the resonance and $E_g=1.2$ eV is the energy of the band gap.
This model is appropriate for MC since the absorption peak at 2.2 eV can become very strong while the film remains transparent at lower energies.

\renewcommand{\thefigure}{S5}
\begin{figure}[!b]
\includegraphics[scale=0.6]{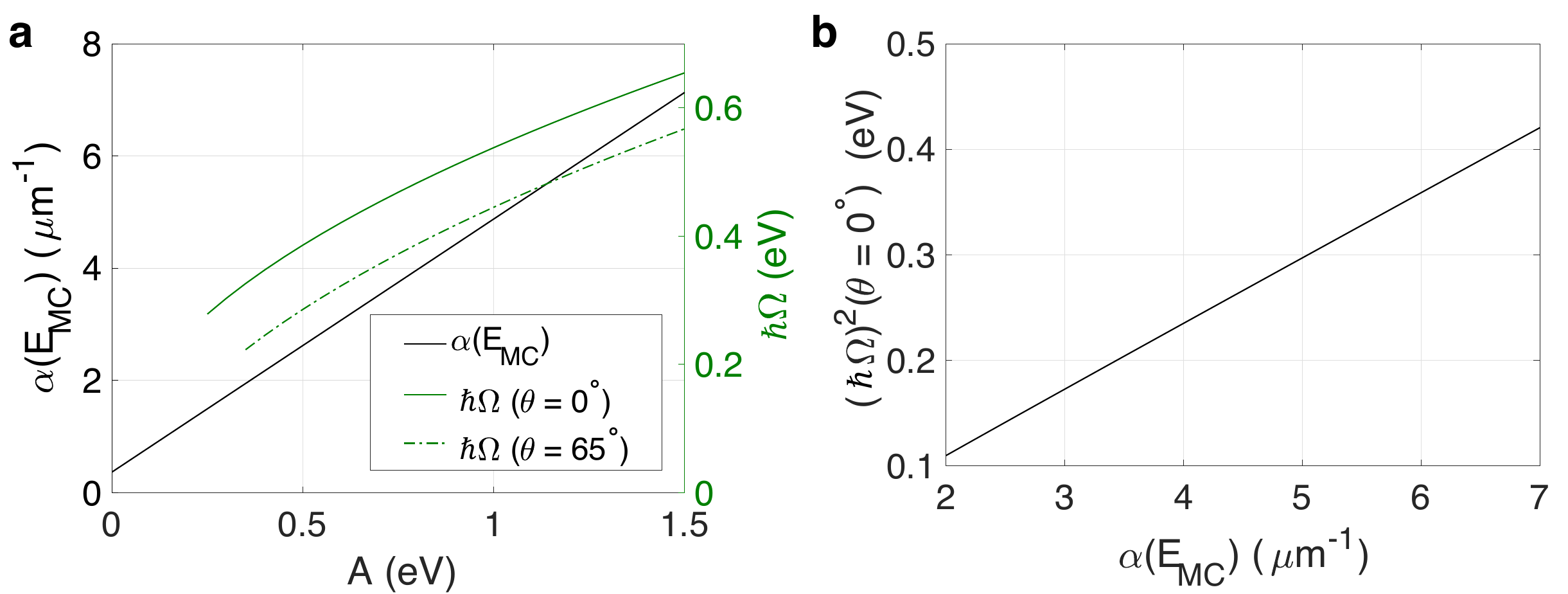}
\centering
\caption{(a) The effect of Tauc-Lorentz oscillator amplitude on $\alpha(E_{\text{MC}})$ (left axis, black line) and $\hbar\Omega$ (right axis, green lines).
(b) Relationship between $\alpha(E_{\text{MC}})$ and $(\hbar\Omega)^2$.}
\label{fig:alphavsamp}
\end{figure}

In Figure \ref{fig:alphavsamp} we plot the relationship between $A$, $\alpha(E_{\text{MC}})$ and the Rabi splitting $\hbar\Omega$.
In panel (a) we see a linear relationship between $\alpha(E_{\text{MC}})$ and $A$, with $\alpha(E_{\text{MC}})$ remaining small but slightly positive at $A=0$ resulting from the tail end of strong absorption peaks in the UV.
As expected, $\hbar\Omega$ increases with increasing $A$.
The electric field strength inside a cavity is greater at $\theta=0^{\circ}$ than at $\theta=65^{\circ}$: hence, strong coupling at $\theta=0^{\circ}$ requires a lower value of $A$ and yields a greater magnitude of splitting than it does at $\theta=65^{\circ}$.
In panel (b) we plot $(\hbar\Omega)^2$ against $\alpha(E_{\text{MC}})$ and observe a linear relationship between the two quantities.
Since $\hbar\Omega \propto \sqrt{N}$ (where $N$ is the number of MC molecules in the film), this indicates that there is a linear relationship between $\alpha(E_{\text{MC}})$ and the number of MC molecules in the film.
We therefore conclude that studying the growth of $\alpha(E_{\text{MC}})$ with time is a suitable way of tracking the growth of MC concentration in the SPI/MC film.

We quantified the rate of conversion in a given structure with a conversion constant $\kappa$, the reciprocal of the time it takes for the SPI/MC film to reach $A=0.70$ eV i.e. $\alpha(E_{\text{MC}}) = 3.5$ µm$^{-1}$ in that structure.
This target value was chosen because it is high enough to place the system well within the strong coupling regime ($\hbar\Omega = 0.36$ eV at $\theta=65^{\circ}$).
For our normal incidence measurements (main article Fig. 4), the reaction proceeded much more rapidly, so a target value of $A=1.2$ eV i.e. $\alpha(E_{\text{MC}}) = 5.8$ µm$^{-1}$ ($\hbar\Omega = 0.59$ eV at $\theta=0^{\circ}$) was chosen to reduce experimental error.
Note, however, that while the choice of target $\alpha(E_{\text{MC}})$ has an impact on the absolute value of $\kappa$, this choice has little effect on the trends in $\kappa$ as film thickness varies (see below).

Each sample was exposed to UV radiation for 1,100 s, with one ellipsometry measurement (consisting of one $\Psi$ spectrum and one $\Delta$ spectrum) taken every 11 s.
This allowed us to observe the conversion of SPI into MC and the transition from the weak to the strong coupling regime.
An example is plotted in main article Fig. 1d-e.
Each vertical strip of main Fig. 1d is a $\Psi$ spectrum acquired at a given time.
First, we determine the cavity thickness by fitting to the ellipsometry spectrum at $t=11$ s.
When determining the cavity thickness we assume that for this initial measurement the MC concentration is negligible, i.e. $A_{\text{MC}}=0$.
We then fix the thickness and set $A_{\text{MC}}$ as the fit parameter for all measurements.

\addcontentsline{toc}{subsection}{S2.2 Comparison of photoisomerisation in strongly coupled and fully detuned cavities}
\subsection*{S2.2 Comparison of photoisomerisation in strongly coupled and fully detuned cavities}

\renewcommand{\thefigure}{S6}
\begin{figure}[!t]
\includegraphics[scale=0.48]{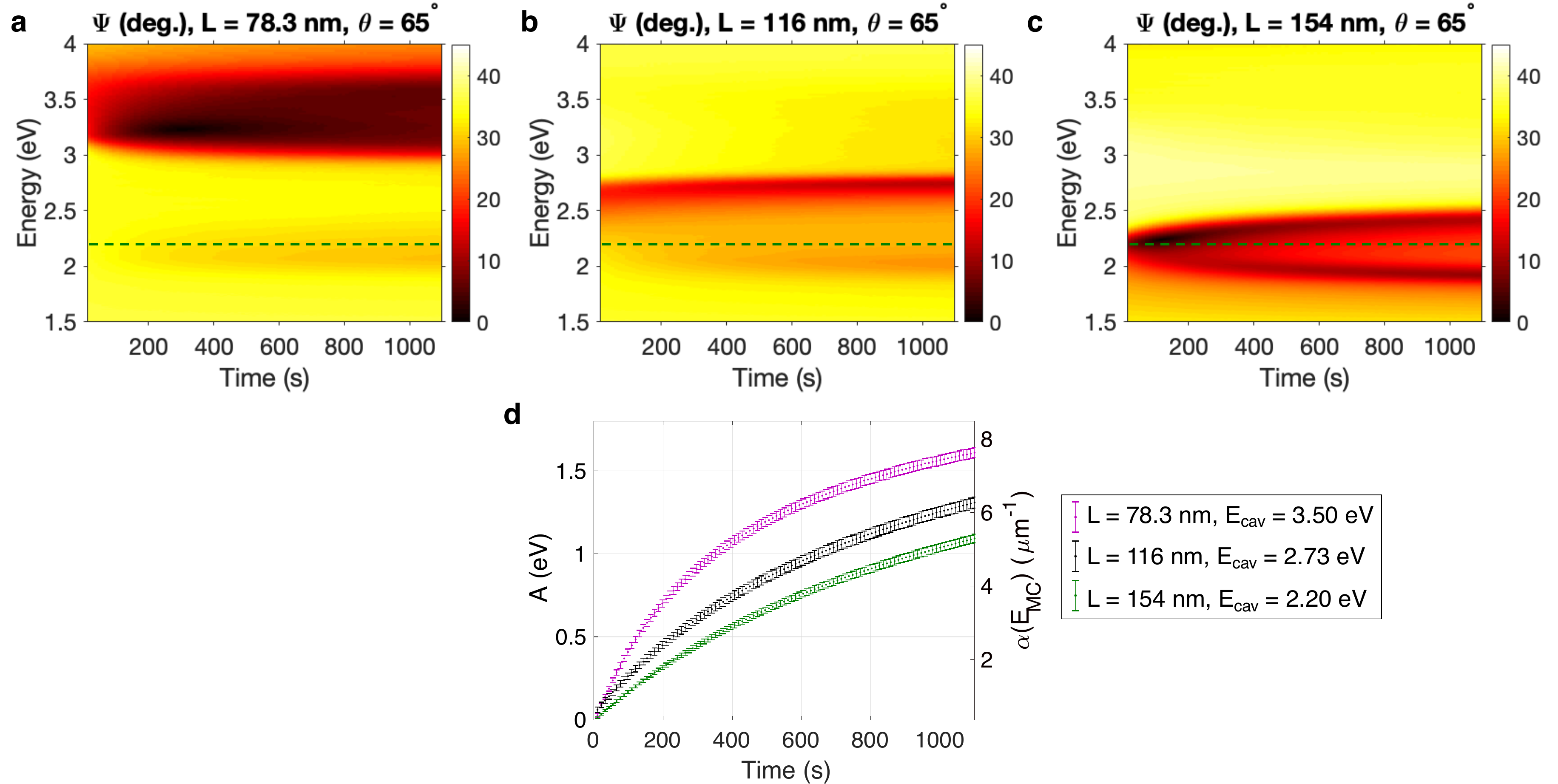}
\centering
\caption{(a-c) Spectra of the ellipsometric parameter $\Psi$ of three Al cavity (thicknesses (a) 78.3 nm, (b) 116 nm and (c) 154 nm) as a function of UV irradiation time acquired at an incident angle of $\theta=65^{\circ}$.
The dashed green lines indicate the energy of the MC's 2.2 eV resonance.
(d) $A$ (left axis) and $\alpha(E_{\text{MC}})$ (right axis) as a function of UV exposure time for the three cavities in panels (a-c): cavity thickness 78.3 nm supporting a cavity mode at 3.50 eV (magenta points), thickness 116 nm supporting a cavity mode at 2.73 eV (black points) and thickness 154 nm supporting a cavity mode at 2.2 eV (green points).}
\label{fig:timept1}
\end{figure}

In Fig. \ref{fig:timept1}a-c we plot the way the $\Psi$ spectra for three Al cavities change as a function of UV exposure time.
All data are acquired at an incident angle of $\theta=65^{\circ}$.
In panel (a) we plot data acquired for a cavity of thickness 78.3 nm, for which the first-order cavity mode occurs at an energy of 3.5 eV. In panel (b) the thickness is 116 nm and the first-order cavity mode occurs at 2.7 eV. In panel (c) the thickness is 154 nm and the first-order cavity mode occurs at 2.2 eV, matching the MC molecular resonance, the energy of which has been marked in all plots by the dashed green line.
As expected, in panel (c) we see the initial uncoupled cavity mode split into two polariton branches.
However, we also see polariton branches form in panels (a) and (b): both plots show a new mode appearing, but since neither of these features are centred on 2.2 eV they must be polariton branches emerging from the strong coupling of MC with the detuned cavity modes.
The observation of strong coupling for such high detuning (as much as 1.3 eV) challenges the idea that one can create a cavity that is fully detuned with no strong coupling; we further explore this idea in Supplementary Section S6.

In Fig. \ref{fig:timept1}d we plot how $A$ and $\alpha(E_{\text{MC}})$ changes with time for these three cavities.
The error bars for $A$ are calculated from the $90\%$ confidence intervals for $A$, which are calculated during fitting to ellipsometry data (performed using the Levenberg-Marquardt method: a standard, iterative, non-linear regression algorithm).
We note that $\alpha(E_{\text{MC}})$ varies with time in the same manner as in Hutchison \textit{et al.}'s work~\cite{hutchison2012modifyingSI}.
The curves are nonlinear and cannot be modelled by a simple linear or exponential relationship.
The reaction rate therefore changes as a function of time.
However, we note from Fig. \ref{fig:timept1}c that for $L=154$ nm the strong coupling regime is reached within the first 200 s.
Yet within this time the gradient of $\alpha(E_{\text{MC}})$ remains approximately constant.
This contrasts with Hutchison \textit{et al.}'s report~\cite{hutchison2012modifyingSI} that entering the strong coupling regime causes a change in reaction rate, which implies we would expect to see a change in gradient before 200 s.

Nevertheless, Fig. \ref{fig:timept1} could easily lead one to the same conclusion as Hutchison \textit{et al.}~\cite{hutchison2012modifyingSI}: the reaction rate increases as the cavity is detuned from the strong coupling condition.
Does the correlation between the suppression of photoisomerisation when the cavity thickness matches the condition for strong coupling have a causal explanation, or is photoisomerisation enhanced when the cavity mode moves into the ultraviolet spectrum?
Based on the data presented in Fig. \ref{fig:timept1} (or, indeed, in Hutchison \textit{et al.}'s work), it is not possible to tell.
We therefore expanded our work to study photoisomerisation trends for a wide range of film thicknesses.

\newpage
\addcontentsline{toc}{subsection}{S2.3 Comparison of photoisomerisation rates before and after entering the strong coupling regime}
\subsection*{S2.3 Comparison of photoisomerisation rates before and after entering the strong coupling regime}

\renewcommand{\thefigure}{S7}
\begin{figure}[!h]
\includegraphics[scale=1.2]{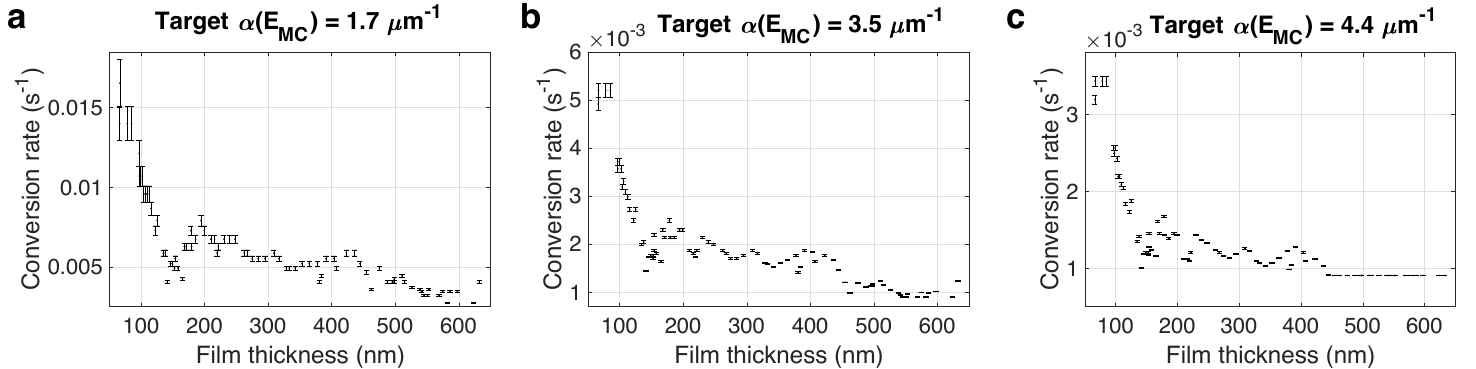}
\centering
\caption{Reaction rate plots for Al cavities probed at $\theta=65^{\circ}$ calculated with different target $\alpha(E_{\text{MC}})$. In (a) we plot the reciprocal of the time taken to reach $\alpha(E_{\text{MC}}) = 1.7$ µm$^{-1}$, in (b) we use a target of $\alpha(E_{\text{MC}}) = 3.5$ µm$^{-1}$ and in (c) we use a target of $\alpha(E_{\text{MC}}) = 4.4$ µm$^{-1}$.}
\label{fig:timept2}
\end{figure}

\noindent To obtain a better picture of what influences the photosisomerisation rates in cavity structures, for the remainder of this work we studied many cavities with a wide range of thicknesses, in all cases sweeping through at least two conditions for strong coupling. (Main Figs. 2-4, Supp Figs. \ref{fig:multiangle} onward.)
Unless otherwise specified we calculated $\kappa$, as defined above, using a target of $A=0.70$ eV i.e. $\alpha(E_{\text{MC}}) = 3.5$ µm$^{-1}$.
For on-resonance structures this corresponds to a splitting of $\hbar\Omega = 0.36$ eV at $\theta=65^{\circ}$.
Here we show that the trends of $\kappa$ with film thickness are approximately independent of our choice of target $\alpha(E_{\text{MC}})$.
In Fig. \ref{fig:timept2} we plot $\kappa$ for the same Al cavity dataset ($\theta=65^{\circ}$) using three different targets for $\kappa$.
In panel (a) we choose $A=0.30$ eV i.e. $\alpha(E_{\text{MC}}) = 1.7$ µm$^{-1}$ which is insufficient for strong coupling; in panel (b) we use $A=0.70$ eV i.e. $\alpha(E_{\text{MC}}) = 3.5$ µm$^{-1}$, which gives $\hbar\Omega = 0.36$ eV (the values that we use for the majority of this study); in panel (c) we use $A=0.90$ eV i.e. $\alpha(E_{\text{MC}}) = 4.4$ µm$^{-1}$, which gives $\hbar\Omega = 0.42$ eV.
All three plots reveal effectively the same trend.

Hutchison \textit{et al.}~\cite{hutchison2012modifyingSI} say that, in samples capable of supporting strong coupling, the reaction rate decreases upon entering the strong coupling regime and decreases further as $\hbar\Omega$ increases.
(However, the authors do not state when their on-resonance cavity enters the strong coupling regime, it is impossible to verify this from the results they present.)
In contrast, our results suggest that trends in reaction rate are not changed when entering the strong coupling regime.

\newpage
\newpage
\addcontentsline{toc}{section}{S3. Photochemistry in Al cavities at a range of incident angles}
\section*{S3. Photochemistry in Al cavities at a range of incident angles}
\renewcommand{\thefigure}{S8}
\begin{figure}[!h]
\includegraphics[scale=1]{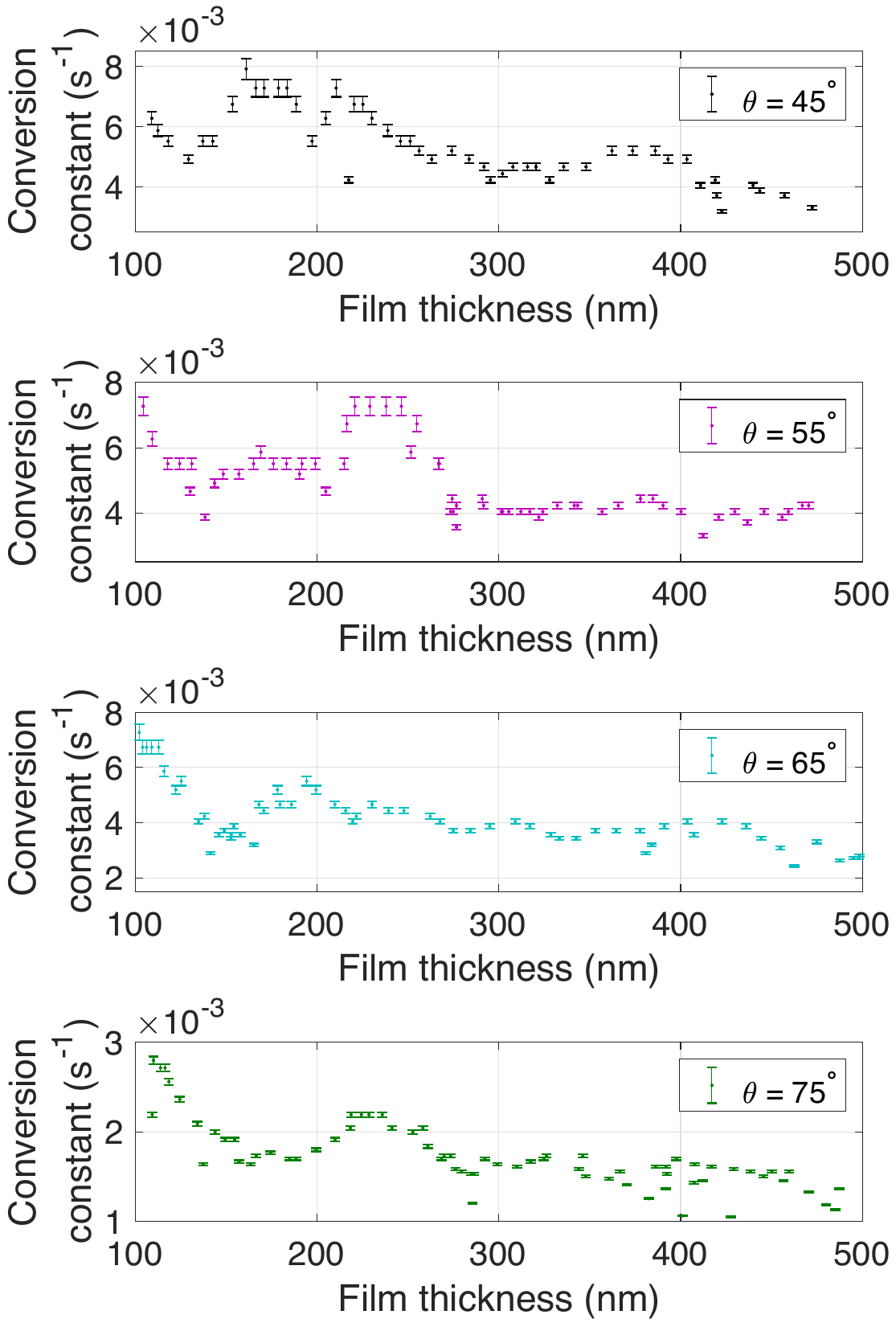}
\centering
\caption{Photoisomerisation conversion constant measured in Al cavities for a range of SPI/MC film thicknesses with incident angles $\theta=45^{\circ}$, $55^{\circ}$, $65^{\circ}$ and $75^{\circ}$.
At $\theta=75^{\circ}$ most films never reach the target of $\alpha(E_{\text{MC}}) = 3.5$ µm$^{-1}$; hence, we reduce the target to $\alpha(E_{\text{MC}}) = 2.2$ µm$^{-1}$ when calculating $\kappa$ at all angles in this figure.
As $\theta$ is increased the magnitude of the conversion constant decreases (since at higher $\theta$  more light is reflected from the top mirror and never enters the cavity) and the peaks and troughs in the curves shift to higher film thicknesses (consistent with the shifting of the absorption peaks at 3.5 eV).
Otherwise, the relationship between conversion constant and film thickness is unchanged.
}
\label{fig:multiangle}
\end{figure}

\newpage
\addcontentsline{toc}{section}{S4. Photochemistry in non-Al structures}
\section*{S4. Photochemistry in non-Al structures}
\renewcommand{\thefigure}{S9}
\begin{figure}[!h]
\includegraphics[scale=1]{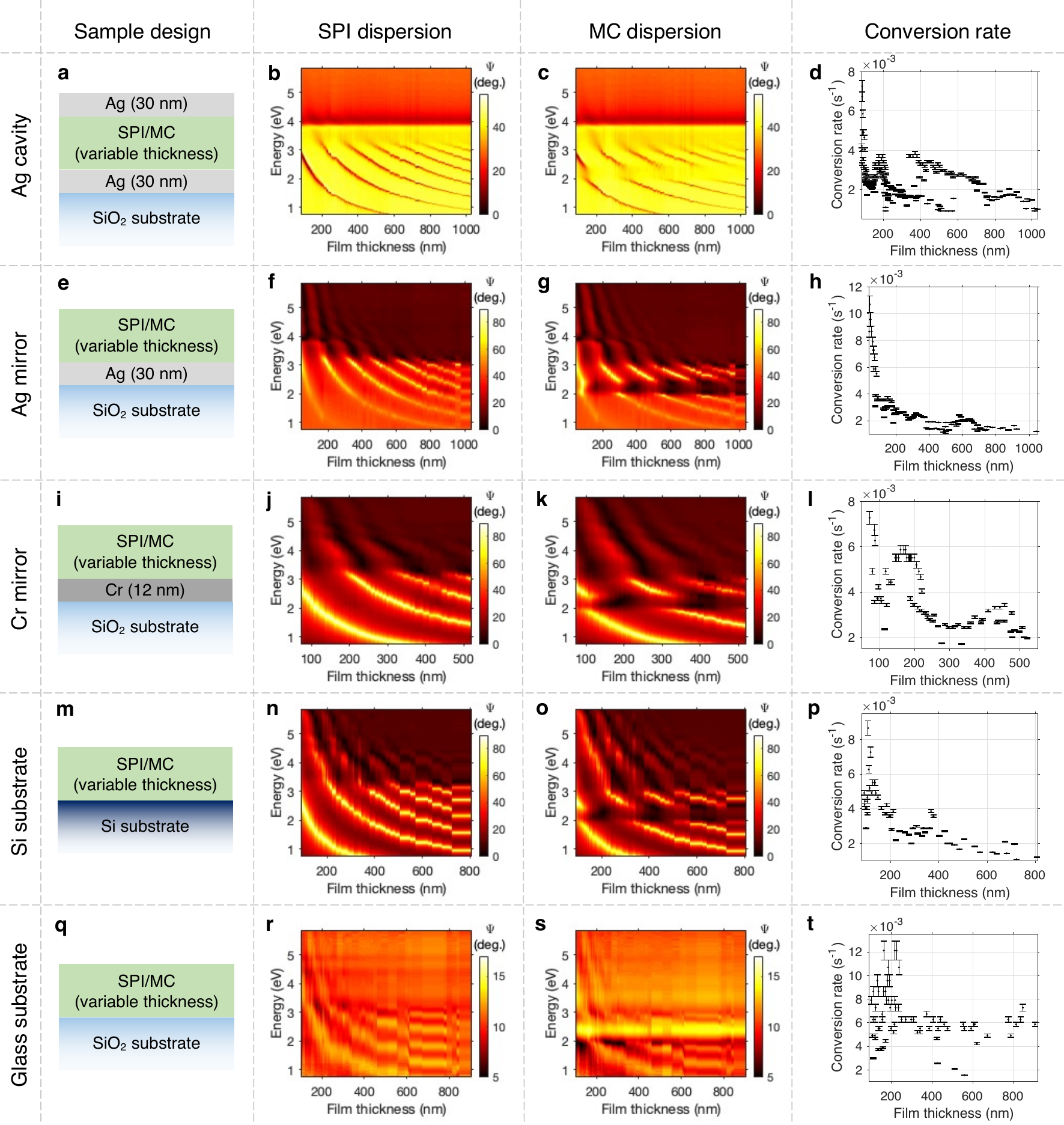}
\centering
\caption{Sample design (a,e,i,m,q), SPI (b,f,j,n,r) and MC (c,g,k,o,s) dispersion plots constructed from $\Psi$ spectra, and SPI conversion rate plots (d,h,l,p,t) for Ag cavity (a-d), Ag mirror (e-h), Cr mirror (i-l), Si substrate (m-p) and amorphous glass substrate (q-t). All data acquired at $\theta=65^{\circ}$.
The target absorption coefficient used to calculate $\kappa$ for the Ag cavity (panel d) was reduced from the usual $\alpha(E_{\text{MC}}) = 3.5$ µm$^{-1}$ to $\alpha(E_{\text{MC}}) = 2.6$ µm$^{-1}$ because the thickest Ag cavity samples never attained $\alpha(E_{\text{MC}}) = 3.5$ µm$^{-1}$. The trend in the plotted data is otherwise the same as when using a target of $\alpha(E_{\text{MC}}) = 3.5$ µm$^{-1}$.
The anticrossing feature at 4 eV in the SPI and MC Ag mirror plots (f,g) is caused by impedance matching between the Ag and SPI/MC film and is unrelated to strong coupling~\cite{tan2022origin}.
}
\end{figure}

\newpage
\addcontentsline{toc}{section}{S5. Accounting for photochemical reaction rates with calculated SPI absorption}
\section*{S5. Accounting for photochemical reaction rates with calculated SPI absorption}

\addcontentsline{toc}{subsection}{S5.1 Calculation of SPI film absorption}
\subsection*{S5.1 Calculation of SPI film absorption}
We have established that photoisomerisation rate is strongly dependent on SPI/MC film thickness in a wide variety of structures.
To try and understand the relationship between $\kappa$ and film thickness, we calculated the absorption of SPI films in these various structures as a function of film thickness.
These calculations, based solely on the optical constants of SPI and other materials (Ag, Al, etc), are  independent of the experimental ellipsometry data used to calculate photosiomerisation rates.

The absorption profile of the SPI layer can be calculated by finding the power loss density within the layer, which can be obtained by finding the divergence of the average Poynting vector within the film and integrate over the thickness of the SPI film. The power loss density is given by
\begin{equation}
\nabla \cdot \vec{\bar{S}} = -2\kappa k_0 |\vec{\bar{S}}|
\end{equation}
where $\vec{\bar{S}}$ is the average Poynting vector, $\kappa$ is the imaginary part of permittivity, and $k_0$ is the magnitude of wavevector in freespace. 
By normalising the power loss density to the energy of the incident beam, the absorption per unit distance within the layer is~\cite{pettersson1999modelling, burkhard2010accounting}:
\begin{equation}
P(z)=\frac{2\kappa k_0 \text{Re}(n_{\text{SPI}} \cos\theta_m)}{\text{Re}(n_1 \cos\theta_1) } F(z)
\end{equation}
where $n_{\text{SPI}}$ is the refractive index of SPI, $\theta_m$ is the refractive angle within the SPI layer, $F(z)=E(z)^2/E_0^2$, $E(z)$ is the electric field within the SPI layer, and $E_0$ is the electric field of the incident beam.
The absorption profile can then be obtained by integrating $P(z)$ over the thickness of the SPI layer. 
The average SPI absorption is then calculated by dividing the absorption profile by the thickness of the SPI layer.

\addcontentsline{toc}{subsection}{S5.2 Correlation between SPI film absorption and photochemical reaction rates}
\subsection*{S5.2 Correlation between SPI film absorption and photochemical reaction rates}

\renewcommand{\thefigure}{S10}
\begin{figure}[!t]
\includegraphics[scale=1.35]{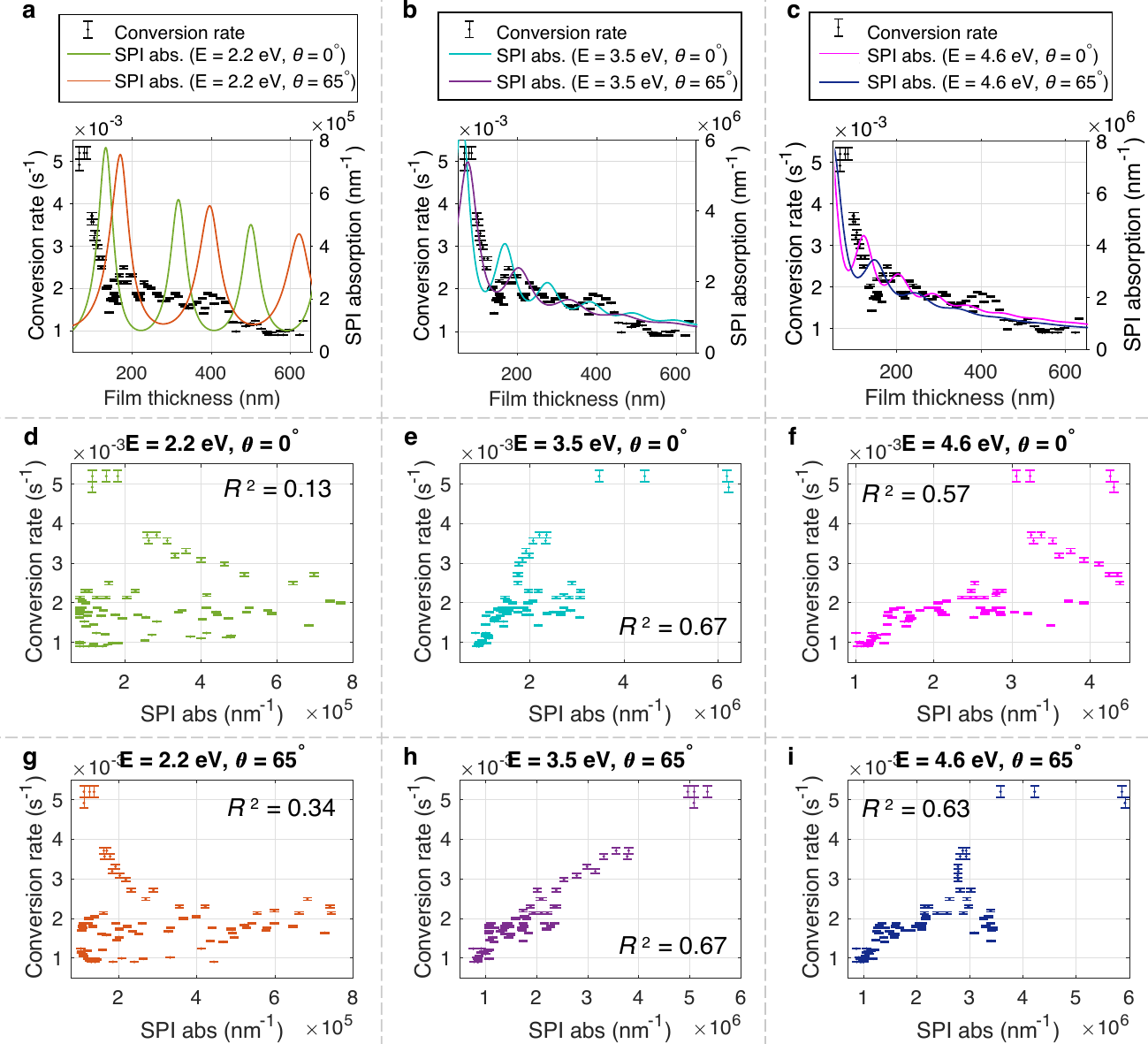}
\centering
\caption{Comparing experimentally determined SPI conversion rates in Al cavities with calculated SPI absorption normalised to SPI film thickness for (a) light of energy 2.2 eV at incident angles $0^{\circ}$ (green line) and $65^{\circ}$ (orange line), (b) energy 3.5 eV at $\theta=0^{\circ}$ (cyan line) and $\theta=65^{\circ}$ (purple line), and (c) energy 4.6 eV at $\theta=0^{\circ}$ (magenta line) and $\theta=65^{\circ}$ (blue line).
In panels (d)-(i) we plot scatter plots to determine the correlation between each of the calculated SPI absorption profiles in panels (a)-(c) and the experimental reaction rate data.}
\label{fig:r2}
\end{figure}

In Figs. \ref{fig:r2}a-b we reproduce main article figs. 3a-b, which compare the experimentally measured reaction rate data with SPI absorption profiles calculated for different energies and incident angles (as outlined above).
In addition to Figs. \ref{fig:r2}a-b, which show SPI absorption profiles calculated for $E=2.2$ eV, 3.5 eV at $\theta=0^{\circ}$, $\theta=65^{\circ}$, in panel (c) we plot the calculated absorption profiles for $E=4.6$ eV (which corresponds to another absorption peak in SPI, see main article Fig. 1b) at $\theta=0^{\circ}$ and $\theta=65^{\circ}$.
The 3.5 eV and 4.6 eV profiles all do a good job of reproducing the trend in reaction rate over hundreds of nanometres, with the 3.5 eV, $\theta=65^{\circ}$ profile giving the best prediction of undulations that happen when the film thickness is varied by a few tens of nanometres.

It is hard to see any clear relationship between either of the 2.2 eV absorption profiles and the experimental data.
In panels d-i we present scatter plots with experimental conversion rates plotted against the various calculated SPI absorption profiles.
In panels (d,g) we see no clear correlation between conversion rate and the 2.2 eV absorption, suggesting there is no measurable effect of strong coupling on reaction rate.
All the UV absorption profiles in panels (e,f,h,i) show a clearer correlation, but it is the $E=3.5$ eV, $\theta=65^{\circ}$ plot that shows the clearest relationship.
Note that while the $E=3.5$ eV, $\theta=0^{\circ}$ and $E=3.5$ eV, $\theta=65^{\circ}$ absorption data give approximately the same $R^2$, it is clear from inspecting panels (e) and (h) that the $E=3.5$ eV, $\theta=65^{\circ}$ data is a much better predictor of conversion rate than $E=3.5$ eV, $\theta=0^{\circ}$.

\subsection*{S5.3 Limitations in our analysis}
Our SPI absorption profiles were calculated using a single discrete energy (e.g. 3.5 eV), when in reality the absorption peaks of interest have a finite width.
All our measurements were made using focussing optics, so that our samples were not only illuminated with light at one discrete angle $\theta$ but over a narrow range of angles centred around $\theta$.
Given this, it is remarkable that the match between our relatively simplistic $E=3.5$ eV, $\theta=65^{\circ}$ SPI absorption calculations and photosiomerisation rate curves is as good as it is, especially at lower thicknesses.
\addcontentsline{toc}{subsection}{S5.3 Limitations in our analysis}

We also note that the $R^2$ values calculated and plotted in main Figure 3c exist as a summary of Supplementary Figure \ref{fig:accounting}, which is too large to include in the main manuscript.
Analysis based solely on $R^2$ values should be treated with caution, since one can compute the same $R^2$ from hugely different data distributions (see e.g. Figs. \ref{fig:r2}e,h,i or Anscombe's quartet).
The main limiting factor on the upper limit of $R^2$ is the level of noise in the experimental data, which varies from sample to sample.
One can reasonably compare the magnitude of $R^2$ values calculated from the same experimental dataset (e.g. to compare the match between 3.5 eV and 2.2 eV at $65^{\circ}$ for the Al cavity dataset). However, it does not make much sense to compare the $R^2$ values for e.g. Al cavity and Ag cavity (since one would expect the Ag cavity dataset to give lower $R^2$ values anyway because it is slightly noisier).

Finally we note that, strictly speaking, our analysis does not definitively prove that strong coupling has no influence on photoisomerisation rate.
We can only say that the trends we do see appear to be well explained by modulation of UV absorption and that  any influence of strong coupling is in comparison negligible.

\newpage
\addcontentsline{toc}{subsection}{S5.4 Conversion rates and SPI absorption at different energies for all structures}
\subsection*{S5.4 Conversion rates and SPI absorption at different energies for all structures}

\begin{figure}[!h]
\includegraphics[scale=1.3]{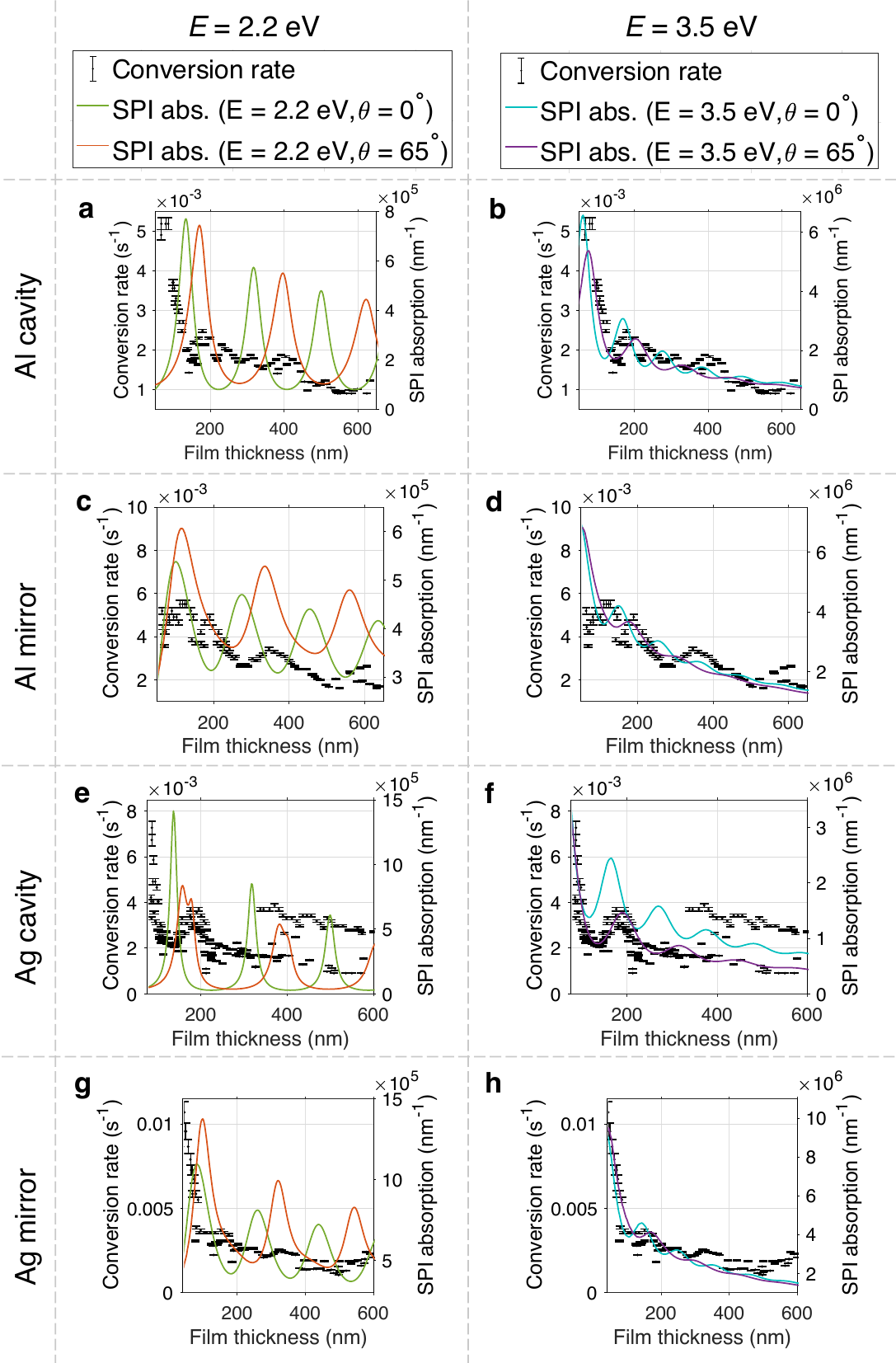}
\centering
\end{figure}
\noindent Figure continues overleaf.

\newpage
\renewcommand{\thefigure}{S11}
\begin{figure}[!h]
\includegraphics[scale=1.3]{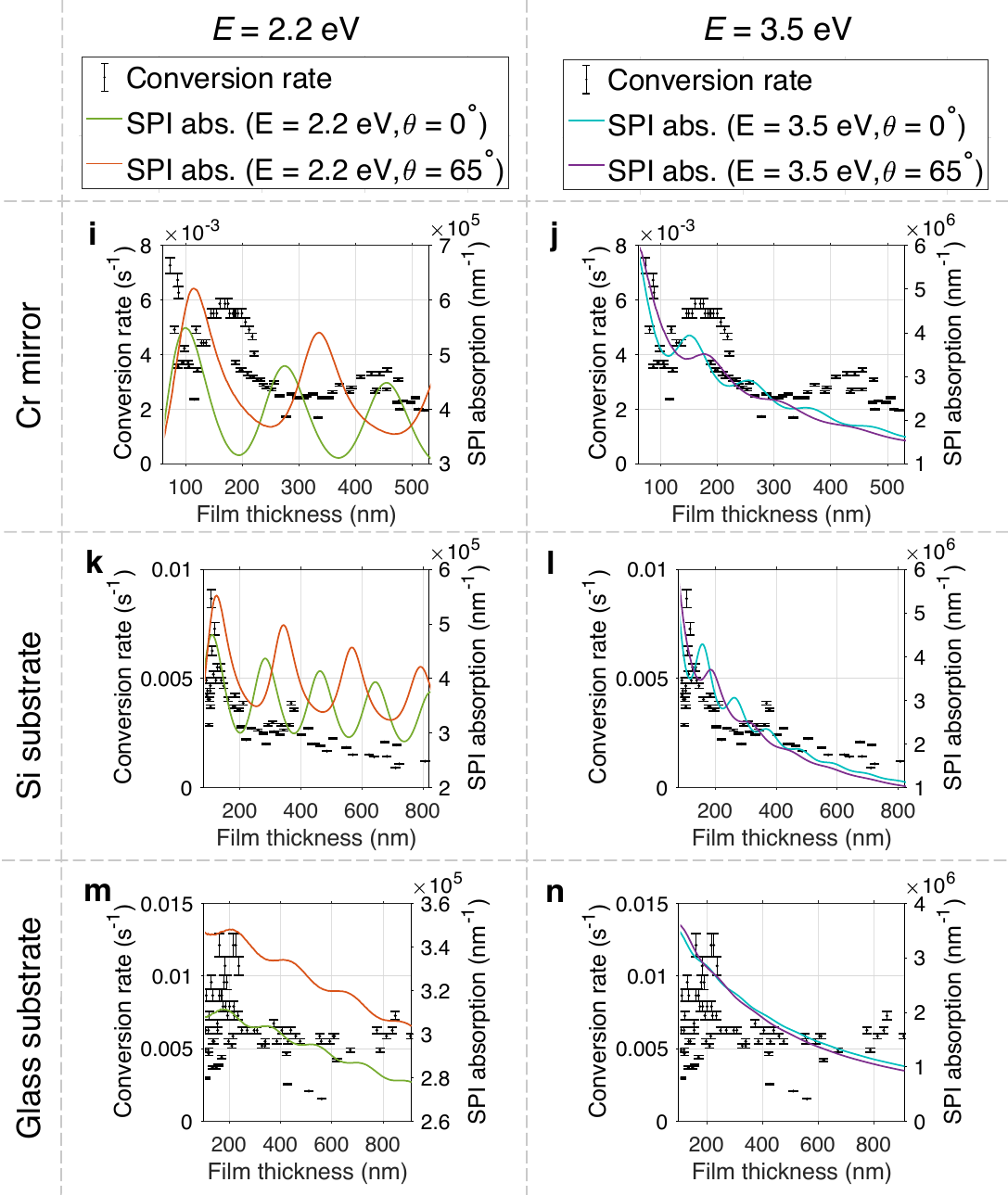}
\centering
\caption{Comparing photoisomerisation rates in a variety of structures with their calculated average absorption at 2.2 eV (left-hand column) and 3.5 eV (right-hand column) at $\theta=0^{\circ}$ and $\theta=65^{\circ}$.
Where there is a clearly resolvable trend in experimental data, the best match (acquired at $\theta=65^{\circ}$) is given by the absorption at 3.5 eV, $\theta=65^{\circ}$.
In no case does the condition for strong coupling (at either $\theta=0^{\circ}$ or $\theta=65^{\circ}$) give a good match to the photoisomerisation rate.
}
\label{fig:accounting}
\end{figure}
The peaks in the $E=2.2$ eV, $\theta=65^{\circ}$ curve for the Ag cavity in panel (e) are slightly split. The exact positions of the cavity modes for $\theta \neq 0^{\circ}$ is polarisation-dependent.
Because our calculations are for unpolarised light, this means that two closely-overlapping modes appear.
This effect is only visible for the Ag cavity data in panel (e) because Ag cavities are the only structures studied whose cavity modes are sufficiently high-quality to clearly reveal this effect.

The noisier nature of our glass substrate dataset (panels m-n) can be explained by the close impedance matching of the SPI/MC film and substrate. This makes the field confinement much weaker, reducing the influence of UV absorption and allowing other sample characteristics (such as roughness) to have a greater influence on conversion rate.

\newpage
\addcontentsline{toc}{section}{S6. Calculated normal incidence transmittance through thin Al/MC/Al cavities}
\section*{S6. Calculated normal incidence transmittance through thin Al/MC/Al cavities}
\renewcommand{\thefigure}{S12}
\begin{figure}[!h]
\includegraphics[scale=0.5]{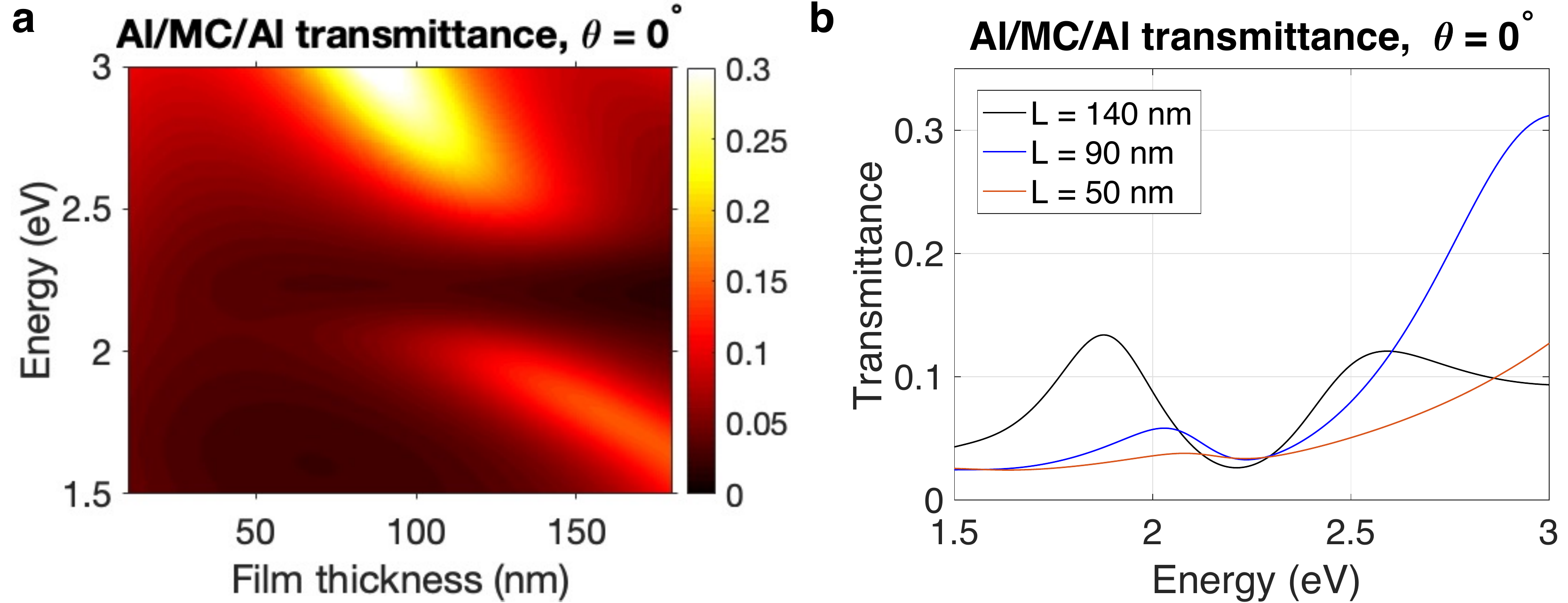}
\centering
\caption{Calculated normal incidence transmittance through quartz (1 mm) / Al(10 nm) / MC($L$) / Al(10 nm) / air structure.
(a) Dispersion plot for a range of MC film thicknesses 10 nm $\leq L \leq$ 180 nm showing clear anticrossing between MC's resonance at 2.2 eV and the first-order cavity mode, whose energy matches the MC resonance energy at $L=140$ nm.
The lower polariton band is clearly distinguishable for MC film thicknesses well below 100 nm and only starts to disappear once the film thickness approaches $L=50$ nm.
In panel (b) we plot the transmittance spectra corresponding to $L=140$ nm (black line), $L=90$ nm (blue line) and $L=50$ nm (orange line), reiterating that the lower polariton band is clearly visible at $L=90$ nm and still faintly visible at $L=50$ nm.
These results are consistent with the $\theta=65^{\circ}$ Al cavity data plots in Supplementary Figure \ref{fig:timept1}, where splitting is also observable even when the first-order cavity mode is highly detuned from 2.2 eV.}
\end{figure}

\noindent In the context of Hutchison \textit{et al.}'s work~\cite{hutchison2012modifyingSI}, these results are significant.
To determine the effect of strong coupling on the SPI/MC photoconversion process, Hutchison \textit{et al.} compared the photoisomerisation rate of SPI in an ``on-resonance'' cavity (thickness $L=130$ nm) to the photoisomerisation rate of SPI in a thinner ``off-resonance'' cavity.
The thickness of this ``off-resonance'' control cavity is unspecified; however, our calculations suggest that, in order to be a meaningful control, the ``off-resonance'' cavity thickness would need to be substantially thinner than 90 nm.
Given that Hutchison \textit{et al.}'s cavities contained two 20 nm PVA spacer layers separating their mirrors from the SPI/MC film, our calculations suggest that the SPI/MC film would have had to be extraordinarily thin (below 50 nm - something we could not achieve in our experiments with the SPI molecule concentration necessary for ultrastrong coupling).

\newpage
\addcontentsline{toc}{section}{S7. Quantum yield as a function of cavity thickness}
\section*{S7. Quantum yield as a function of cavity thickness}
\renewcommand{\thefigure}{S13}
\begin{figure}[!h]
\includegraphics[scale=0.5]{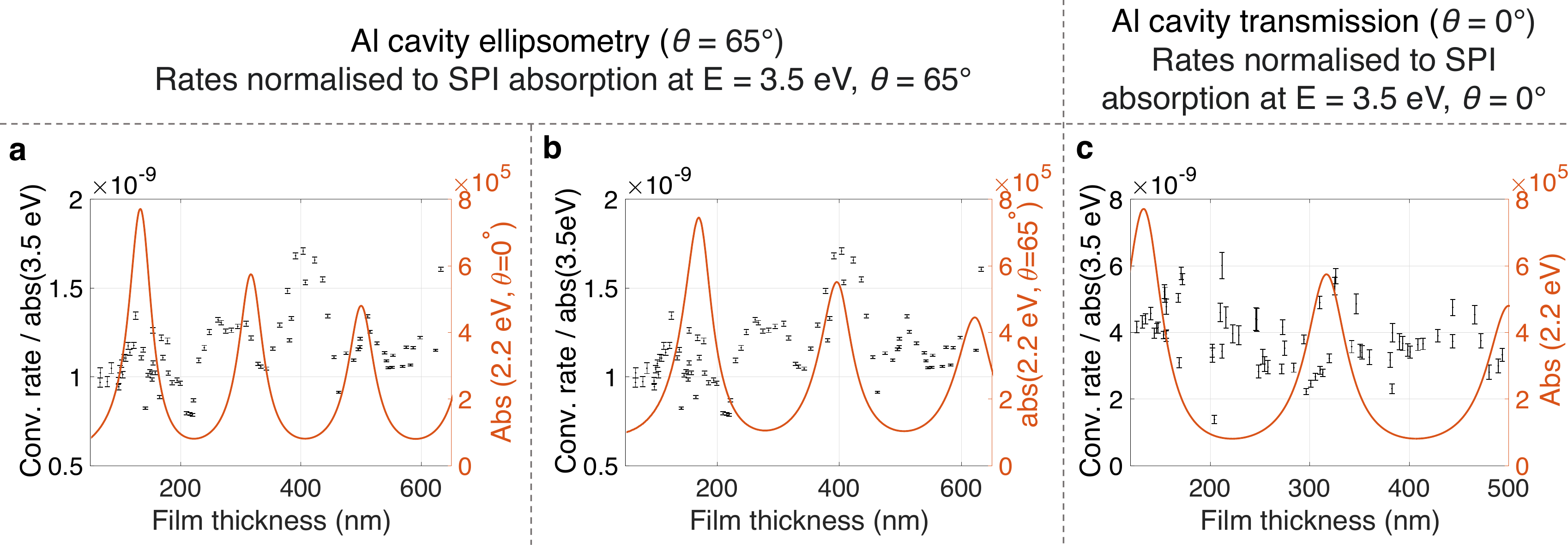}
\centering
\caption{Black points: Reaction rate divided by SPI's absorption of UV radiation ($E=3.5$ eV) for (a,b) ellipsometry experiment at $\theta=65^{\circ}$ (Main Figs. 2-3) and (c) normal incidence transmission experiment (Main Fig. 4).
Orange curves: Calculated SPI absorption at 2.2 eV for (a) $\theta=0^{\circ}$ (strong coupling at normal incidence), (b) $\theta=65^{\circ}$ (strong coupling at $\theta=65^{\circ}$) and (c) $\theta=0^{\circ}$ (strong coupling at normal incidence).}
\end{figure}

\noindent
Since the rate of a photochemical reaction is the rate of photons absorbed by the reactant times the quantum yield of the reaction (i.e. the number of molecules that complete the reaction per absorbed photon), our results suggest that strong light-matter coupling with MC does not affect the quantum yield of the SPI-MC photoisomerisation process.
Here we plot our reaction rate data normalised to SPI absorption at 3.5 eV, which is proportional to the quantum yield of the SPI-MC process.
We also plot the SPI absorption curves for 2.2 eV, whose maxima indicate the cavity thicknesses required for strong coupling.
We observe no correlation between reaction quantum yield and the conditions for strong coupling.

\end{document}